\documentclass[acmsmall,screen]{acmart}

\usepackage{algorithm}
\usepackage{algpseudocode}
\usepackage{listings}
\usepackage{enumitem}
\usepackage{tikz}
\usepackage{xcolor}
\usepackage{verbatim}
\usepackage{multirow}
\usepackage{graphicx}
\usepackage{wrapfig}
\usepackage{tcolorbox}
\usepackage{tabularx}
\usepackage{subcaption}
\usepackage{amsmath}

\AtBeginDocument{%
  \providecommand\BibTeX{{%
    \normalfont B\kern-0.5em{\scshape i\kern-0.25em b}\kern-0.8em\TeX}}}

\setcopyright{acmcopyright}
\copyrightyear{2025}
\acmYear{2025}
\acmDOI{XXXXXXX.XXXXXXX}

\acmJournal{TOSEM}
\acmVolume{0}
\acmNumber{0}
\acmArticle{1}
\acmMonth{0}

\begin{document}

\title{ACTesting: \underline{A}utomated \underline{C}ross-modal \underline{Testing} Method of Text-to-Image Software}

\author{Siqi Gu}
\email{siqi.gu@smail.nju.edu.cn}
\orcid{0000-0001-5514-6734}
\affiliation{
  \institution{State Key Laboratory for Novel Software Technology, Nanjing University}
  \city{Nanjing}
  \state{Jiangsu}
  \country{China}
  \postcode{210093}
}

\author{Chunrong Fang}
\email{fangchunrong@nju.edu.cn}
\authornotemark[1]
\orcid{0000-0002-9930-7111}
\affiliation{
  \institution{State Key Laboratory for Novel Software Technology, Nanjing University}
  \city{Nanjing}
  \state{Jiangsu}
  \country{China}
  \postcode{210093}
}

\author{Quanjun Zhang}
\email{quanjun.zhang@smail.nju.edu.cn}
\orcid{0000-0002-2495-3805}
\affiliation{
  \institution{State Key Laboratory for Novel Software Technology, Nanjing University}
  \city{Nanjing}
  \state{Jiangsu}
  \country{China}
  \postcode{210093}
}

\author{Zhenyu Chen}
\email{zychen@nju.edu.cn}
\orcid{0000-0002-9592-7022}
\affiliation{
  \institution{State Key Laboratory for Novel Software Technology, Nanjing University}
  \city{Nanjing}
  \state{Jiangsu}
  \country{China}
  \postcode{210093}
}

\renewcommand{\shortauthors}{S. Gu et al.}
\begin{abstract}
Recently, creative generative artificial intelligence software has emerged as a pivotal assistant, enabling users to generate content and seek inspiration rapidly. Text-to-Image (T2I) software, one of the most widely used, synthesizes images with text input by engaging in a cross-modal process. However, despite substantial advancements in the T2I engine, T2I software still encounters errors when generating complex or non-realistic scenes, including omitting focal entities, low image realism, and mismatched text-image information. The cross-modal nature of T2I software complicates error detection for traditional testing methods, and the absence of test oracles further exacerbates the complexity of the testing process. To fill this gap, we propose \textbf{ACTesting}, an \textbf{A}utomated \textbf{C}ross-modal \textbf{Testing} Method of Text-to-Image Software, the first testing method explicitly designed for T2I software. ACTesting utilizes the metamorphic testing principle to address the oracle problem and identifies cross-modal semantic consistency as its fundamental Metamorphic relation (MR) by employing the Entity-relationship (ER) triples. We design three kinds of mutation operators under the guidance of MR and the adaptability density constraint to construct the new input text. After generating the images based on the text, ACTesting verifies whether MR is satisfied by detecting the ER triples across two modalities to detect the errors of T2I software. In our experiments across five popular T2I software, ACTesting effectively generates error-revealing tests, resulting in a decrease in text-image consistency by up to 20\% when compared to the baseline. Additionally, an ablation study demonstrates the efficacy of the proposed mutation operators. The experimental results validate that ACTesting can reliably identify errors within T2I software.

\end{abstract}



\maketitle

\section{Introduction}
\begin{figure}[!tbp]
  \centering
  \includegraphics[scale=0.50]{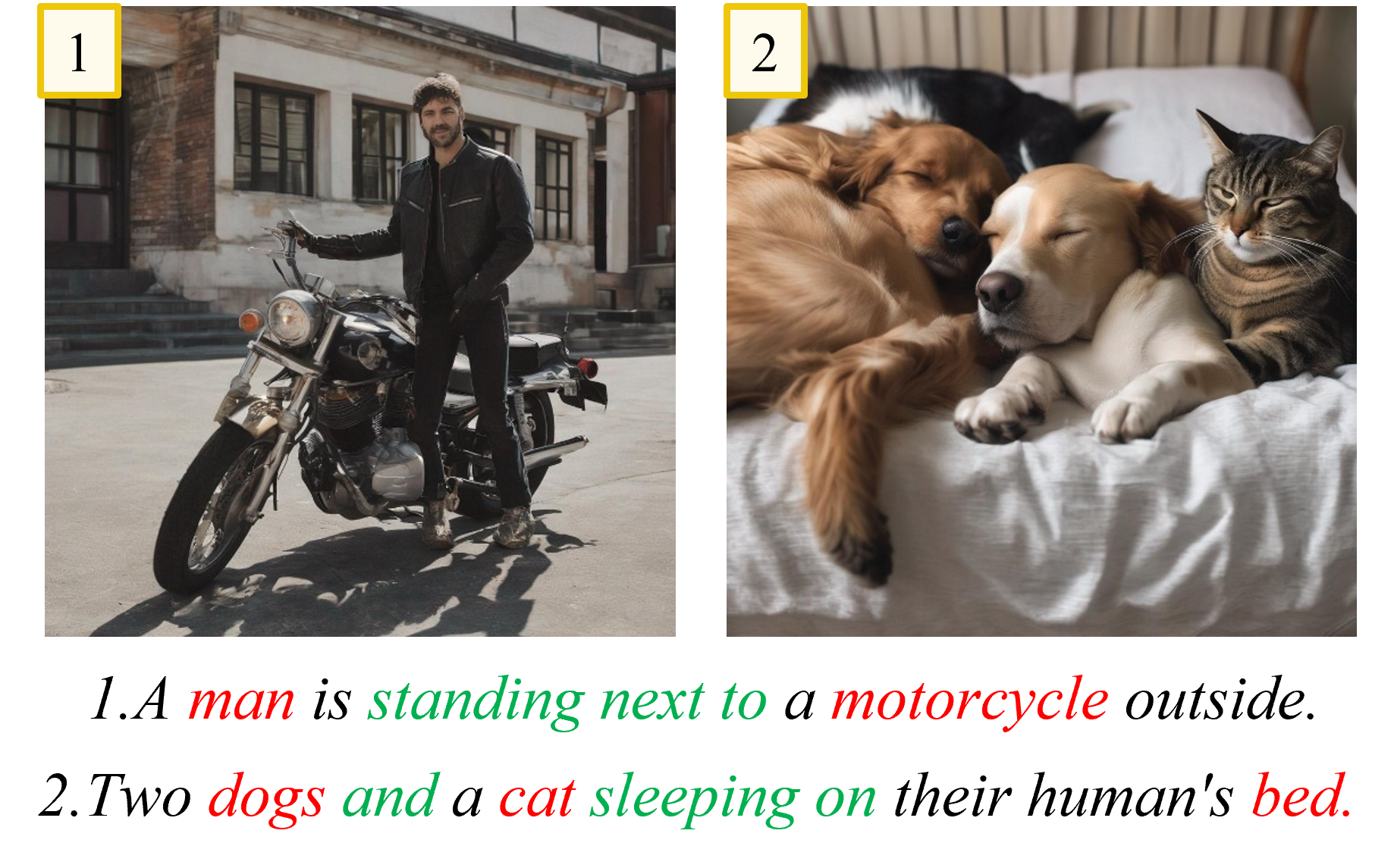}
  \caption{Sample inputs and outputs of T2I software.}
  \label{backgroud}
\end{figure}

Deep learning has undergone significant evolution. The rise in popularity of transformer~\cite{devlin2018bert}, generative adversarial network (GAN)~\cite{zhu2017unpaired} and diffusion model~\cite{rombach2022} empower Artificial Intelligence Generative Content (AIGC) to yield surprising results in fields, such as image, text, and audio generation within simple inputs. The emergence of large-scale models precipitates a rise in creative generative tasks. Among them, Text-to-Image (T2I) task is one of the most popular ones, aiming to automatically synthesize a creative image based on the simple input text as shown in Fig.~\ref{backgroud}. Recently, several T2I models~\cite{zhang2017stackgan, huang2019realistic, lao2019dual, Gu2022, Saharia2022, qiao2019mirrorgan, rombach2022, kim2022verse} and software~\cite{baidu, WanXiang, DALLE, Stability, DeepAI, Midjourney} from top IT companies show great performance on fidelity and creativity of the output image. The high-quality content and the simple interaction flow make increasing use of T2I software in our daily lives. Representative applications include generating the image for cross-modal data augmentation~\cite{dataaugmentation} (Computer Science), artistic creation and designing~\cite{Art1} (Art), and multilingual communication assistant for deaf~\cite{medicial} (Medical). The ability to handle cross-modal information and understand rich scenarios greatly enhances T2I software's potential for development.

However, the outputs from T2I software are not entirely reliable or capable of meeting the expected requirements. Despite diverse techniques that have been researched and adapted to improve the inner engine of T2I software, the generative results could be abnormal or incorrect~\cite{hinz2020semantic, 9879249}, especially under unreal-world scenarios. The reasons are the huge and complex neural network structure, labeling errors in training datasets, and uneven feature distribution. Especially, the existing evaluation metrics are unexplainable (mainly based on deep feature similarity) and cannot accurately pinpoint the errors in the generated images. Generating unpleasant, offensive, or inappropriate image content may result in significant repercussions~\cite{dangers}. Objectionable or discriminatory content can damage a brand in a business setting and even trigger public protests or legal action. Consequently, it is significant to test the generation quality of T2I software.

To the best of our knowledge, however, there is no testing method specifically designed for T2I software available. Developing testing methods for T2I software presents persistent challenges ~\cite{frolov2021adversarial, BORJI201941, theis2015note}. Initially, in contrast to conventional software with delineated internal logic, commercial T2I software kernels are more complex, including both deep learning engines and software modules. Most of the end-users can only get results through the API. This makes white-box testing methods no longer suitable for this task. Secondly, the most intricate issue is cross-modal alignment, necessitating aligning two modes with completely different information representations during the T2I software's testing phase. Thirdly, the simple testing methods based on text augmentation, such as swapping the order of the words or introducing spelling errors, possibly cause errors in semantic information before generation. In addition, current research~\cite{lee2024holistic} focuses on quantifying aspects such as image realism or alignment through deep learning models, but they do not explicitly identify defects within the images. Therefore, testers cannot precisely determine which semantic representations in the generated images are problematic.

To tackle the above-mentioned challenges, we propose ACTesting, an automated and black-box cross-modal testing method for T2I software. ACTesting is designed based on the Metamorphic Testing (MT) theory to address the issue of oracle scarcity. \textbf{The core design concept of Metamorphic relation (MR) in ACTesting is the cross-modal semantic information matching consistency. That is, the semantic information changes of the input text and the output image should be consistent}. To construct the MR, we employ the Entity-relationship (ER) model. Specifically, to preserve semantic preservation, we apply the ER triple to represent the focal semantic information in both modalities different from traditional text augmentation. We also propose three mutation operators based on the adaptability density constraint instead of the probability-proportional sampling to construct the new input text. Finally, ACTesting constructs input text based on the proposed mutation operators and treats the MR as practical adequacy criteria to reveal hidden errors of tested T2I software. During the implementation process, ACTesting uses Object Detection and Scene Graph models to estimate the entities and relationships in the generated images, thus eliminating the need for pre-labeling.

To evaluate the effectiveness of ACTesting, we conduct experiments on five widely used T2I software. ACTesting generates 142,170 synthetic images based on the MS-COCO validation dataset~\cite{dinh2022tise} in total to test the performance of T2I software. We employ Improved-Fréchet Inception Distance (I-FID), Improved-Inception Score (I-IS), and R-Precision (RP)~\cite{heusel2017gans, salimans2016improved, xu2018attngan, dinh2022tise} as the basic evaluation metrics for image realism and text-image relevance. We also design the error detection ($Error_e$ and $Error_r$) and miss-detection ($Miss_e$ and $Miss_r$) rates as the metrics, based on the designed MR ~\cite{Yu2022} and the scene graph generation model~\cite{tang2020unbiased} to further identify the specific defects within the images. The experimental results demonstrate that ACTesting can improve the detection performance of errors. ACTesting degrades image quality by 2.9\% to 15\% according to the I-IS metric and reduces text-image match consistency by 7.5\% to 21.1\% according to the RP metric. The average error rate of the three mutation operators is around 60\%, which is 1.75 times higher than the baseline. These results show that ACTesting not only constructs the mutation text samples to reduce the generation quality of the software but also detects specific errors in the generated images. Moreover, we conduct an ablation study to further elucidate the effectiveness of each mutation operator. The results show that the combined mutation operators effectively increase the miss-detection rate beyond that of the single mutation operator. The main contributions of this paper are as follows:
\begin{itemize}[leftmargin=1em]
\item \textbf{Method}. We introduce the \textit{first} automated cross-modal testing method for T2I software based on MT, termed ACTesting. It keeps the core idea of cross-modal semantic preservation and employs the ER triple to construct the MR. We design three kinds of mutation operators under the guidance of adaptability density constraint to detect the errors in the generated images and test the generation quality of T2I software.
\item \textbf{Tool}. We integrate ACTesting into a Python tool. It represents the first black-box automated testing tool for T2I. We make the code and test data of the entire process publicly available at \url{https://github.com/sikygu/ACTesting}.
\item \textbf{Study}. We conduct a comprehensive experiment to evaluate the performance of ACTesting on five industrial T2I software. The results illustrate that our testing method can adeptly produce error-revealing test cases, leveraging our adaptability density-guided operators. The ablation study further shows that the proposed mutation operators can be flexibly combined for enhanced effectiveness.
\end{itemize}

\section{Preliminaries and Motivation}
In this section, we begin by outlining the workflow of modern T2I software. We then detail the T2I models employed by current T2I software, introducing how they effectively convert text into images. Finally, we present the motivation behind our proposal, which sets the stage for a deeper exploration of our innovative contributions to testing T2I software. 

\subsection{T2I Software Workflow}

\begin{figure}[!tbp]
  \centering
  \includegraphics[scale=0.36]{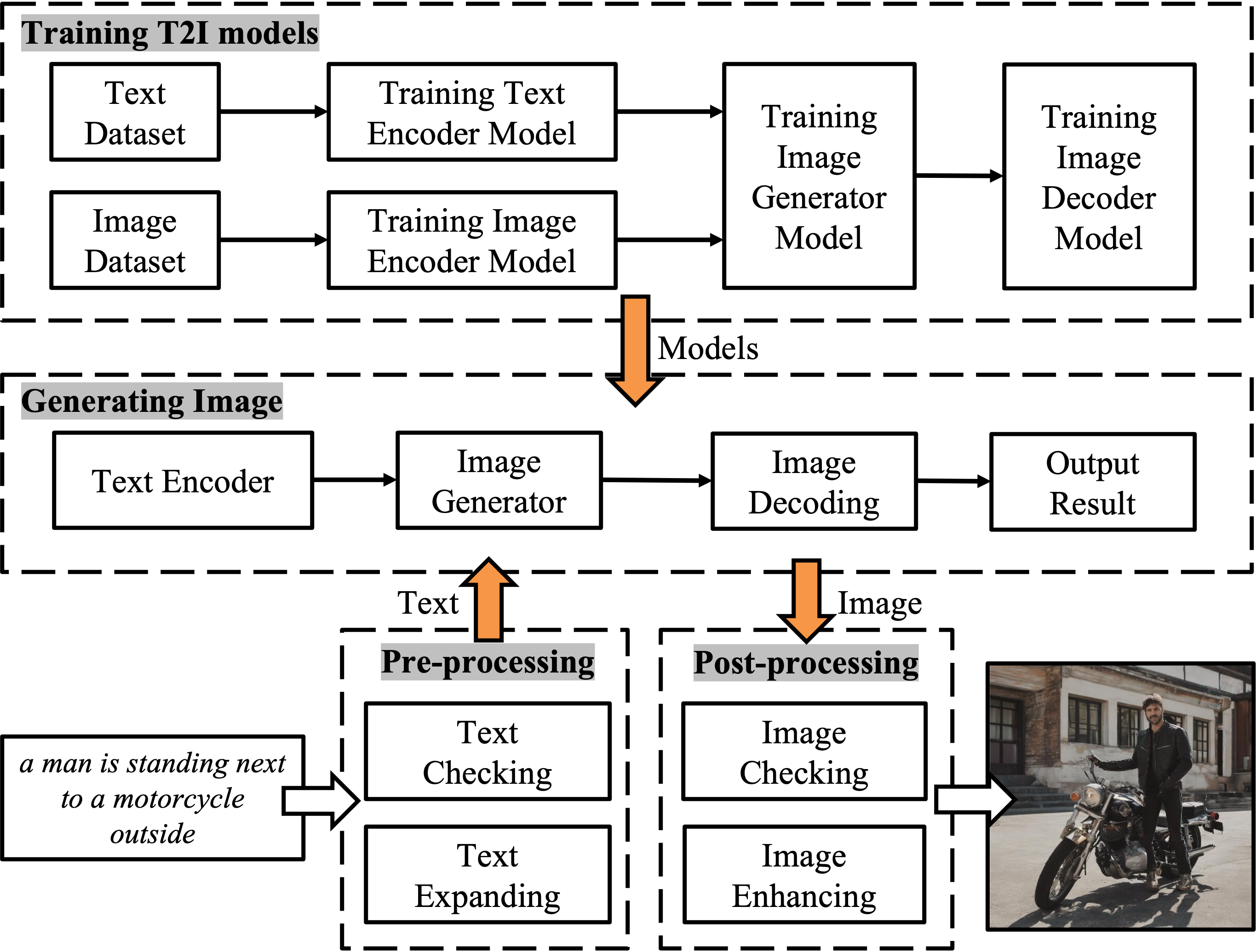}
  \caption{The workflow of modern T2I software.}
  \label{workflow}
\end{figure}

Fig.~\ref{workflow} delineates the general pipeline of modern T2I software, including training T2I models, generating images, pre-processing, and post-processing. Leveraging expansive training image-text datasets, text encoder and image encoder models undergo pre-training to extract the cross-modal feature and establish mappings within a latent space. After that, the image generator and decoder models are trained end-to-end for image information reconstitution. T2I software then deploys these trained models (the text encoder, image generator, and image decoder) for the image synthesis procedure. For more flexible adjustment of input and output, T2I software designs pre-processing and post-processing stages to verify data quality. The pre-processing stage not only ensures text input conforms to predefined standards but also augments more details to the input text to fit the trained models. The initial image is generated based on the processed text input. In the terminal phase, post-processing assesses the initial image for both fidelity and ethical considerations, further employing image enhancement algorithms to mitigate noise and fulfill details. The core part of the whole workflow is the trained T2I models. Thus, we briefly introduce some modern engines in the next section.

\subsection{Modern T2I Models}
With the rapid development of T2I models, two main series of T2I models are proposed: Autoregressive and Diffusion-based T2I models. We introduce cutting-edge methods respectively.

\textbf{Autoregressive Models.} Autoregressive methods can exploit large-scale datasets through time-series models predicated on the dependence of the historical time-series of forecasting targets in different periods. Representative method DALL-E~\cite{ramesh2021zero} based on CLIP~\cite{radford2021learning} from OpenAI regards both text and image tokens as sequential data, performing autoregression via Transformer architecture. Another model, Parti~\cite{yu2022scaling} from Google, encodes images as sequences of discrete tokens and reconstructs these sequences into high-quality images. 

\textbf{Diffusion-Based Models.} The most predominant approach is diffusion-based methods, which are also the new state-of-the-art models in T2I generation. Diffusion models (DMs) aim to reserve a process of perturbing the data with noise for sample generation. As a milestone work, Stable diffusion~\cite{rombach2022} trains the diffusion models within latent space, incorporating the text tokens during the denoising phase to fuse the cross-modal features. Predicated Diffusion~\cite{sueyoshi2024predicated} is proposed recently, a unified framework designed to more effectively express users’ intentions. Predominantly, contemporary T2I software engines are grounded in diffusion models. 

In the experiments, we conduct a comprehensive evaluation by testing five different text-to-image (T2I) software programs. These programs are carefully selected to represent the two distinct types of engines that are widely used in the field. Through this approach, we aim to demonstrate the versatility and broad applicability of ACTesting across diverse T2I generation techniques. By including multiple software platforms with varying architectures and capabilities, we ensure a robust examination of our method's effectiveness and its potential to adapt to different underlying technologies.

\subsection{Motivation}
\label{Motivation}

\begin{figure}[!tbp]
  \centering
  \includegraphics[scale=0.50]{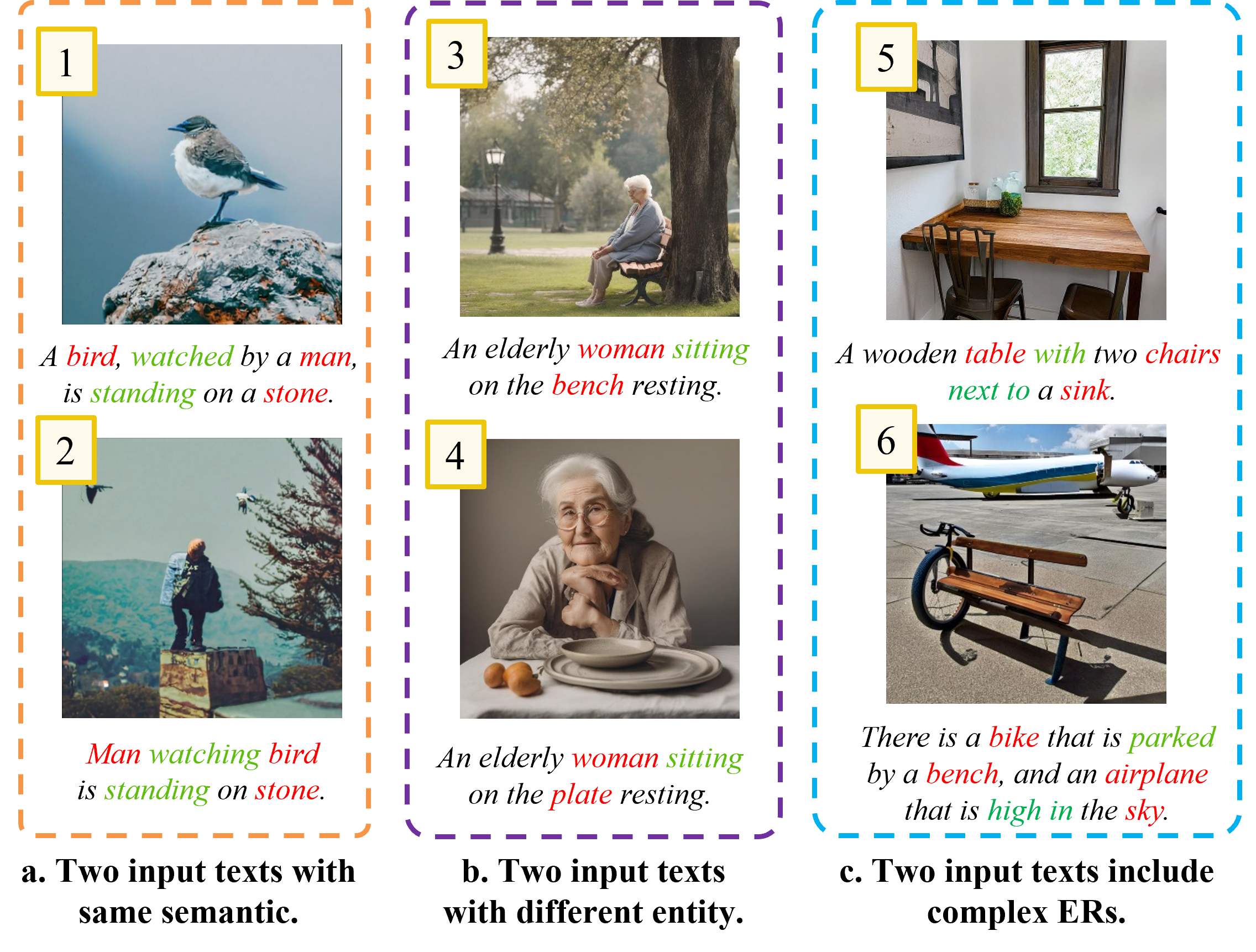}
  \caption{Examples of the input-output pairs of text-to-image software. The images are generated with OpenAI.DALLE and Stable Diffusion XL.}
  \label{motivation}
\end{figure}

The motivation of ACTesting is described from the perspective of cross-modal semantics. Fig.~\ref{motivation} shows images generated by prevalent T2I software highlighting the entities in red and the relationships in green. In Fig.~\ref{motivation}~(a), two sentence variations with the same semantics are as inputs. Although the input text conveys nearly identical meanings, Image~1 omits the ``man'' while Image~2 includes the correct entities and relationships. In Fig.~\ref{motivation}~(b), the common collocation in the sentence ``woman sitting on the bench'' is replaced with ``woman sitting on the plate''. The generated images demonstrate that collocating unreal-world entity relationships degrades the quality of the generated entity (distortion of hand depiction in Image~4). Moreover, in Fig.~\ref{motivation}~(c), complex entities and relationships diminish the realism of the images.

In conclusion, T2I software exhibits high sensitivity to the focal ERs, where even simple alterations in sentence structure could lead to decreased text-image consistency and image quality. This stimulates us to consider that the cross-modal semantic consistency can be used as the core MR of MT in ACTesting. As a result, ACTesting introduces the ER model and using ER triples to be the basic unit of MR.

\section{Approach}

\begin{figure*}[!tbp]
  \centering
  \includegraphics[width=0.75\linewidth]{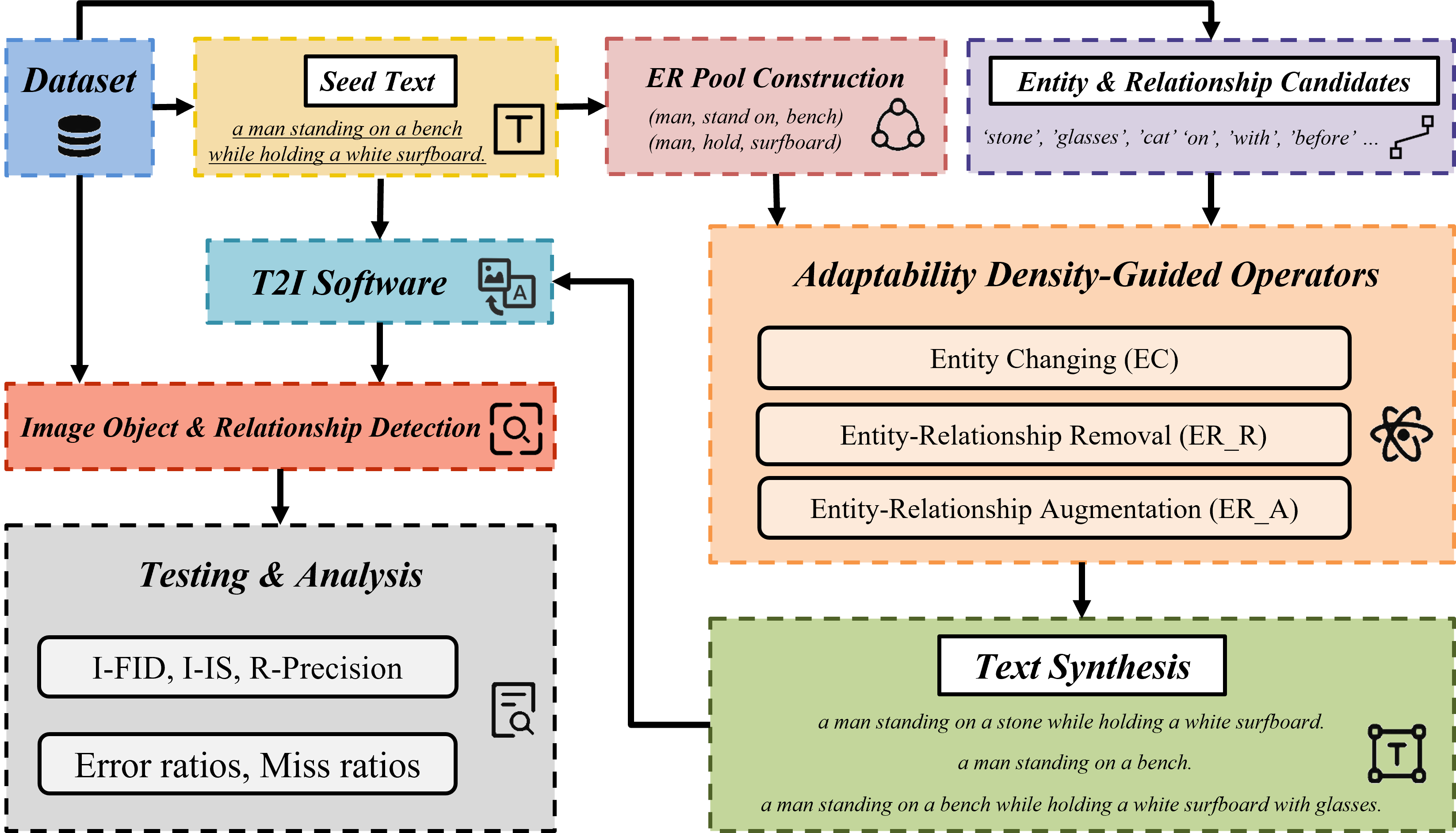}
  \caption{The overview of ACTesting.}
  \label{overview}
\end{figure*}

This section introduces ACTesting, which is proposed to test T2I software automatically and detect the errors of the software in a black-box scenario. ACTesting is based on MT and uses cross-modal semantic consistency as the basic MR. Sec~\ref{Actesting} introduces the workflow of ACTesting according to Fig.~\ref{overview}. The core concept of ACTesting lies in the adaptability density constraint of T2I software, which quantifies its ability to handle diverse test samples and serves as a measure of generation quality (introduced in Sec~\ref{Adaptability}). Then, we characterize the adaptability density constraint by the ER model and define the basic MR in Sec~\ref{MR}. To construct the new input text, we design three mutation operators based on the ER model in Sec~\ref{MO}.

\subsection{Overview of ACTesting}
\label{Actesting}
ACTesting is used to construct a complete testing process. As shown in Fig.~\ref{overview}, ACTesting includes several main processing components: ER pool construction, adaptability density-guided operators, text synthesizing, image entity and relationship detection, and testing and evaluation results calculation. We select seed texts from the dataset that contain diverse ERs (according to the class label of the dataset) at the beginning. Then, ACTesting constructs an ER pool for each seed text based on the principle of semantic information preservation.

\textbf{ER Pool Construction.} The input text in the T2I software is different from other natural language processing tasks like text classification or text abstract. The text often contains the scene or the objects of the generated image, which renders traditional text mutation operators (such as swapping the order of the words or introducing spelling errors in the text) no longer applicable to inputs for this task. Considering the different ways various modalities express information, we transform each input text into an ER pool, utilizing consistent cross-modal semantic information features. Subsequently, all mutation operations are performed based on the ER pool, ensuring that the semantic features of the input text remain intact. We adopt the relationship extraction, entity recognition and structured representation technique to construct the ER pool referring~\cite{qi2020stanza}. This method also constructs the ER candidates based on the open-source dataset.  

\textbf{Mutation Operator Implementation.} To construct the new testing samples and reserve the cross-modal semantic information, three kinds of mutation operators are designed under the adaptability density-guided constraint (The details of the adaptability density-guided constraint and mutation operators are introduced in Sec~\ref{Adaptability} and Sec~\ref{MO}, respectively). ACTesting adopts the EC and ER\_A operators to all the inputs and adopts the ER\_R operator to the input having more than one ER. The mutation operators of ACTesting are shown in Fig.~\ref{mutation}. 

\textbf{Text Synthesizing.} After completing the mutation operations based on the adaptability density-guided constraint, ACTesting constructs the synthesized text based on the mutated ER pools. To ensure the fluency of synthesized sentences and the integrity of semantic information, we utilize BERT~\cite{devlin2018bert} to enhance the details of the sentences. At this point, we obtain paired text inputs. 

\textbf{Image object and relationship detection.} The paired text inputs are sent to the tested T2I software respectively for generating the output images. To address the challenge of cross-modal semantic matching, we employ the object detection model to identify entities in images and use the scene graph model to detect relationships between entities in images during the implementation process. In the end, we forward the detected results to the next component.

\textbf{Testing and Analysis.} Based on the MR and mutation operators introduced in Sec~\ref{MR} and~\ref{MO}, ACTesting computes the test results. We introduce the error-detection ratios and miss-detection ratios for entities and relationships to demonstrate the testing efficiency, detailed in Sec~\ref{metrics}.

In summary, ACTesting encompasses a comprehensive process of new input text generation, image generation by T2I software and test result computation, where each module plays a crucial role in the overall effectiveness of the testing.

\subsection{Adaptability Density-guided Constraint}
\label{Adaptability}

As we mentioned in Sec~\ref{Motivation}, changes in the focal ERs in the input text can easily affect the generated results of T2I software. This means that not all words in the input text impact the quality of the results generated by T2I software. Therefore, traditional random text mutation operators are not effective in thoroughly testing the errors in T2I software. To address this problem, we explore the adaptability density-guided constraint to guide the mutation operators to generate the new input text effectively.

We first formalize the testing process for the T2I software to better understand our proposal. In the T2I software testing phase, one test can be characterized by a set of scenarios $M$ which represent specific input configurations, such as individual text inputs. Each test scenario ${\mu\in M}$ corresponds to a distinct configuration and is evaluated using a performance measure $R$. The performance of software $\pi$ under a given test scenario $\mu$ can be denoted as $R(\pi,\mu)$. To account for variations caused by non-deterministic behaviors or random factors, the expected performance $E[R(\pi,\mu)]$ is utilized, representing the average outcome over multiple test runs for scenario $\mu$. Given the complexity of T2I software structures, different scenarios often yield varying performance results. The overall performance of the software across all scenarios in $M$ is quantified using the aggregated metric $\Phi\left(\pi,M,p\right)$. This metric incorporates $p(\mu)$, which denotes the probability or normalized weight of executing scenario $\mu$, to reflect its relative importance. The aggregated performance metric is formally defined as:

\begin{equation}
\Phi\left(\pi,M,p\right)=\sum_{\mu\in M}\Phi\left(\pi,\mu,p\right)\ =\sum_{\mu\in M}{p(\mu)\bullet}\mathbb{E}[R(\pi,\mu)]
\label{1}
\end{equation}

T2I software $\pi$ is large and complicated enough which cannot be tested across the entire set $M$. Random sampling guided by the distribution $p$ is often employed. However, when $M$ and $p$ are used to define the testing benchmark, probability-proportional sampling based on $p$ may not always provide the most effective testing strategy. Because the set $M$ can include different problems simultaneously (e.g., text understanding, image generation, and image composition), each of which may require different testing priorities. $R$ can also be nondeterministic and/or subject to measurement error. Therefore, the probability-proportional sampling can be inefficient and costly.

Given the inefficiencies and potential costs of probability-proportional sampling, a more purposeful approach to sampling is necessary to effectively approximate Eq.~\ref{1}. ACTesting focuses on testing the generation quality of T2I software, which is defined as the software's ability to perform correctly across diverse scenarios. To achieve this, we aim to assess the software's adaptability density—its capacity to handle varied test inputs while maintaining functionality. For instance, as shown in Fig.~\ref{motivation}, Image~2 illustrates the higher adaptability density compared with Image~1 because it successfully synthesizes all the entities (``man'', ``bird'', ``stone'') and relationships (``watching'', ``standing on'').

To optimize the sampling method in the testing phase, we introduce the adaptability density constraint \( d: M \to \mathbb{R}^+ \). This constraint quantifies the adaptability density of each test scenario \(\mu \in M\). Ideally, for any T2I software \(\pi\) being tested, the following relationship holds:  
\[
\Phi\left(\pi, \mu_1, p\right) > \Phi\left(\pi, \mu_2, p\right) \quad \text{if} \quad d\left(\mu_1\right) < d\left(\mu_2\right).
\]  

In other words, higher adaptability density \( d(\mu) \) corresponds to improved evaluation performance. Therefore, we propose the adaptability density-driven sampling method, formulated as:  
\[
\Phi\left(\pi, M, h\right) = \sum_{\mu \in M} \Phi\left(\pi, \mu, h\right) = \sum_{\mu \in M} h(\mu) \cdot p(\mu \mid h(\mu)) \cdot \mathbb{E}[R(\pi, \mu)],
\]  
where \( h(\mu) \) denotes the sampling probability of scenario \(\mu\), weighted by its adaptability density. Under the T2I task, \( h \) is discretized to better account for varying adaptability density levels. To refine the adaptability density constraint, we assert that for any T2I software \(\pi\) being tested and for two adaptability density levels \( a \) and \( b \) (\( a \leq b \)), the following relationship is satisfied:  
\[
D(\Phi\left(\pi, M_a, h\right),\Phi\left(\pi, M_b, h\right))< \epsilon,
\]  
where $D(\cdot,\cdot)$ is a distance metric applied to the evaluation results $\Phi$ aggregated over the respective scenario subsets ($M_a$ \( M_a = \{\mu \mid \mu \in M, d(\mu) = a\} \)) and $M_b$ (\( M_b = \{\mu \mid \mu \in M, d(\mu) = b\} \)). $\epsilon$ is a parameter of constraint. This ensures that scenarios with lower adaptability density contribute more significantly to the evaluation, promoting a robust and balanced assessment of the software’s performance.

Typically, $\pi$ remains a black box to end-users, obscuring the exact techniques it employs. However, it is unequivocally necessary for this testing task to align text and images, implying a requirement for consistency in entities and their interrelationships, albeit represented differently. Given these considerations, we specify the adaptability density constraint corresponding to the matching degree of semantic information in two modalities.

However, it is difficult to calculate the adaptability density of T2I software directly based on the semantic information. The contrastive language-image pre-training (CLIP)~\cite{radford2021learning} model is unsuitable, as it is highly vulnerable to attacks and lacks sufficient interpretability to provide further explanations to testers for anomalies. Therefore, to better represent semantic information in cross-modal data, and given that the input text for T2I software is mostly scene descriptions, which is highly analogous to scene understanding in the field of computer vision, we adopt the ER model to represent semantic information.

Specifically, as shown in Fig.~\ref{backgroud}, we deliver two output images corresponding to two input texts. The salient entities in each image are consistent with the input texts, marked in red (e.g., ``man'', ``motorcycle'', ``cat''). If the input text describes the relationship between focal entities, marked in green (e.g., ``standing next to'', ``and'', ``sleep on''), the generated image should be described as well. Each entity and relationship has its \textit{class}. Based on the entity and relation extraction technique in nature language processing, each text can be constructed to an entity-relation pool (ER pool), containing several entity-relation triples (e.g.,  (``dog'', ``with'', ``cat''), (``dog'', ``on'', ``bed''), (``cat'', ``on'', ``bed'')). All elements in the ER pool are converted to the class they belong to (e.g., ``and'' convert to ``with'', ``sleep on'' convert to ``on'', ``man'' convert to ``person'') referring to~\cite{rvlbert, yu2023visually}. Note that the analysis of singular and plural nouns is not the focus of our proposal.

\subsection{Metamorphic Relation}
\label{MR}

One of the central challenges in testing T2I software is the oracle problem—the difficulty of objectively determining whether the output matches the target. This challenge arises from the inherent subjectivity in assessing complex outputs like images, where the correspondence between input prompts and generated visuals cannot be easily quantified. To address these limitations, we adopt MT and define the concept of MR to mitigate potential issues arising from the lack of test oracles. As we analyze in Sec~\ref{Adaptability}, the focal entities and relationships contained in the input text and output image are the primary semantic representations. Therefore, ACTesting uses the ER model to represent the cross-model semantic and takes semantic consistency as the basic MR.

Formally, considering the T2I software as $\pi$, the input text in the seed set $\mathbb{S}=\{s_{i}\}$ ($s_{i}$ represents the $i$-th input text). The adaptability density-guided constraint in ACTesting is defined as $d$. $f_{er}$ is used to represent the cross-modal semantic information. The result obtained after the output image generated by the T2I software based on the input $s_i$ is characterized by the ER model can be represented as:
\begin{equation}
\label{eq2}
    r_{s_i} = {f_{er}}(\pi,s_i)
\end{equation}
Then, we construct new text ($\mathbb{A}=\{a_{j}\}$ ($a_{j}$ represents the $j$-th input text)) based on the adaptability density constraint $d$. Similarly, we use Eq~\ref{eq2} to obtain the result characterized by the ER model:
\begin{equation}
    r_{a_j} = {f_{er}}(\pi,f_m(s_i))
\end{equation}

After obtaining the two outputs, we apply the MR to determine whether the generation requirements of the T2I software are met. Finally, the MR to test the T2I software with the newly generated sample can be formalized as follows:
\begin{equation}
    \forall s_i \in \mathbb{S}, \ \forall a_j \in \mathbb{A}, D(r_{s_i}, r_{a_j})< \epsilon \
\end{equation}
where $\epsilon$ represents the parameter of the constraint.

\begin{figure}[!tbp]
  \centering
  \includegraphics[width=0.48\textwidth]{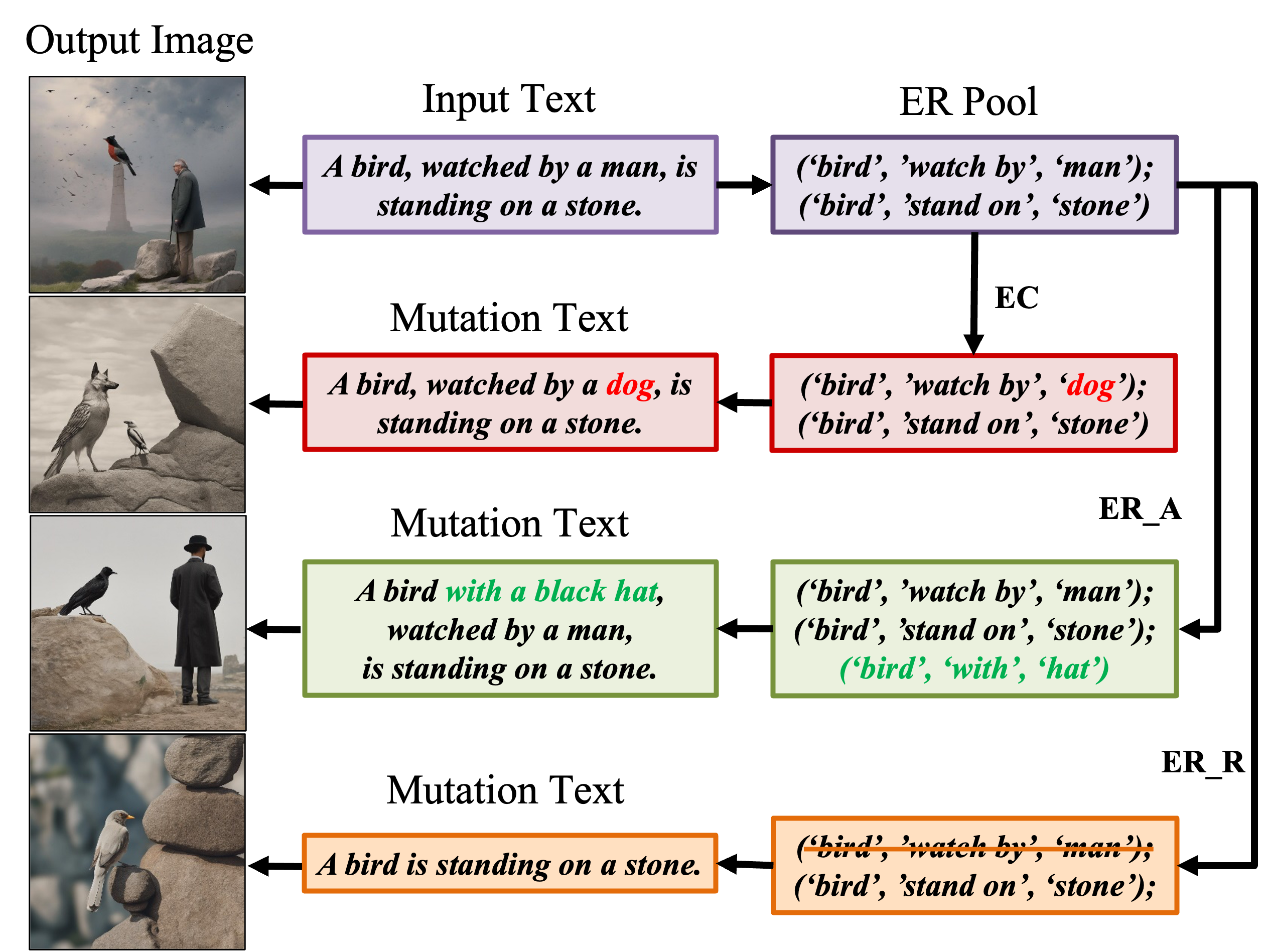}
  \caption{The examples of applying mutation operators of ACTesting.}
  \label{mutation}
\end{figure}

\subsection{Mutation Operators}
\label{MO}

Based on the context of Sec~\ref{Adaptability} and Sec~\ref{MR}, it is crucial to generate new input samples guided by the adaptability density constraint. To this end, we introduce three types of mutation operators designed around the proposed adaptability density constraint. In the construction of text inputs, entities are central elements and are more likely to trigger generation errors in T2I software. Relationships, on the other hand, often serve as subordinate or linking components for entities. Therefore, when designing mutation operators guided by the adaptability density constraint, we define separate principles for entity changing and the unified augmentation or removal of ERs. This ensures semantic validity and preserves the integrity of core content. Fig.~\ref{mutation} illustrates an example of the implementation of these mutation operators in ACTesting. To specify the MR based on cross-modal semantic consistency for T2I software, we refer to ~\cite{Yu2022} and define the MR and corresponding mutation operators as follows:

\begin{equation}
    K_e = Set_e(r_{s_i}) \cap Set_e(r_{a_i}) 
\end{equation}
\begin{equation}
    K_r = Set_r(r_{s_i}) \cap Set_r(r_{a_i})  
\end{equation}
where $Set_e(r_{s_i})$ and $Set_e(r_{a_i})$ denote the set of entity class in $r_{s_i}$ and $r_{a_i}$, $Set_r(r_{s_i})$ and $Set_r(r_{a_i})$ denote the set of relationship class in $r_{s_i}$ and $r_{a_i}$.

\textbf{Entity Changing} (EC): The EC operator substitutes one entity in the ER pool, thus constructing a pair of highly similar input text samples. The output image should ideally remain consistent, aside from the substituted entity after generation. EC operator is primarily employed to ascertain whether T2I software demonstrates variability in generating content for diverse entities. In essence, the EC operator assesses the adaptability of T2I software, examining whether it excessively relies on pre-existing ER in the training dataset, consequently resulting in constrained generation capabilities. EC operator focuses on the issue of generation stability within the context of testing. MR detailed for the EC operator:
\begin{align}
    K_e == Set_e&(r_{s_i}) - Set_e(e_1), K_e == Set_e(r_{a_i}) -Set_e(e_2) \\
    &K_r == Set_r(r_{s_i}), K_r == Set_r(r_{a_i})
\end{align}

\textbf{Entity-relationship Removal} (ER\_R): The ER\_R operator randomly removes an ER triple. Removing one ER triple does not lead to unreadable sentences or semantic errors, which sets it apart from traditional text mutation strategies. The generation of images should ensure that the retained entities and relationships remain unaffected and maintain consistent image quality. ER\_R operator aims at constructing a pair of text samples with differing levels of complexity, used to investigate whether the T2I models exhibit significant disparities in generating these different samples, which is also the reflection of the aspect of generating quality. MR detailed for the ER\_R operator:
\begin{align}
    &K_e == Set_e(r_{a_i}) \\
    &K_r == Set_r(r_{a_i})
\end{align}

\textbf{Entity-relationship Augmentation} (ER\_A): The ER\_A operator randomly adds an entity and its corresponding relationship from candidate entities and relationships to the original ER pool. We construct the candidate ERs from the dataset. After obtaining the new ER pool, we use it as a basis to rebuild the input text. By increasing the variation of the original text input, ACTesting can explore the search space more extensively to uncover potential issues. ER\_A operator aims to increase the complexity of the original samples, expanding the input space to test the generation quality of T2I software. MR detailed for the ER\_A operator:
\begin{align}
    &K_e == Set_e(r_{s_i}) \\
    &K_r == Set_r(r_{s_i})
\end{align}

If any of the $K_e$ or $K_r$ in MR are violated under a certain operator, the generated results will be separately reported as either entity errors ($p_e$) or relationship errors ($p_r$). In summary, the design of the three mutation operators aims to detect errors in T2I software based on MR that maintain consistency in cross-modal semantic information.

\section{Experiment Design}
Our evaluation is designed to answer the three main research questions in the experiments:

\textbf{RQ1}: Can ACTesting generate low-quality output for T2I software? 

\textbf{RQ2}: Can ACTesting generate error-revealing tests for T2I software?

\textbf{RQ3}: How does the combination of different mutation operators impact the generation quality of the tested T2I software?
\subsection{Datasets and T2I Software}
In this section, we introduce the dataset, software and baseline we used in our experiments. 

\textbf{Dataset.} We use the MS-COCO~\cite{caesar2018coco} as the seed sets, which is a widely recognized benchmark in the field of T2I. Importantly, the dataset stands out for its image captioning component, where each image is accompanied by at least five different captions provided by human annotators. These features make MS-COCO highly valuable at the intersection of vision and natural language processing. 

\textbf{T2I Software.} To elucidate the effectiveness of the testing method we propose, we choose five widely-used T2I software using different models in the experiment. We briefly introduce the tested software as follows:

\textit{OpenJourney}~\cite{Openjourney} is one of the most popular text-to-image software. OpenJourney is painting waves in ai-generated art. It is often quoted as a free alternative to MidJourney as it is a Stable Diffusion model trained using thousands of Midjourney images from its v4 update. 

\textit{Wan Xiang}~\cite{WanXiang} is an evolving Artificial Intelligence (AI) painting model ~\cite{huang2023composer} that can create corresponding images or artworks from textual descriptions using machine learning and natural language processing technologies. 

\textit{Stable Diffusion XL}~\cite{Stability} helps users create descriptive images with shorter prompts. The model is a significant advancement in image generation capabilities, offering enhanced image composition and face generation that results in stunning visuals and realistic aesthetics. 

\textit{DeepAI Image Generator}~\cite{DeepAI} creates an image from scratch basedon a text description. It can be used to generate AI art. It provides the functions of AI image generation API call services for developers.

\textit{DALLE2}~\cite{DALLE} is an AI system that can create realistic images from a description in natural language. We choose DALLE2 because it is a more stable version of this series.

\textbf{Baseline.}
ACTesting is the first black-box testing method for T2I software so there are no available baselines. To demonstrate the effectiveness of our testing methodology, we select a commonly used text mutation operator, random synonym substitution (SS), as our baseline. The reason is that SS does not bring unknown semantic errors in the text compared with swapping word positions or introducing spelling errors.

\subsection{Experimental Setup}
\textbf{Implementing details.} We conduct experiments on 4062 seeds of text-image pairs from the MS-COCO validation dataset. A total of six additional test sets were generated across all experiments based on the seeds. We test 150 categories of entities and 50 kinds of relationships based on the dataset setting and labels. \textbf{In summary, we get 5 software * 4062 captions * 7 test sets = 142,170 synthetic images.} All images are resized to 512 $\times$ 512. We calculate evaluation metrics based on all inputs and outputs and obtain the results. To the best of our knowledge, this is the largest experimental evaluation for T2I software.

\textbf{Running environment.} All experiments are performed on a Ubuntu 20.04.6 LTS server with two RTX 3090 GPUs. We implement ACTesting on Python 3.7 and Pytorch 1.10.

\subsection{Evaluation Metrics}
\label{metrics}
To concretize this measurement in the T2I task, we apply two metrics. Firstly, we adopt the I-FID, I-IS, and RP based on~\cite{heusel2017gans},~\cite{salimans2016improved},~\cite{xu2018attngan} and~\cite{dinh2022tise} for evaluating the image realism and text-image relevance. To compute the image realism metrics (I-FID and I-IS), we initially extract features using a pre-trained Inception-v3 network~\cite{szegedy2016rethinking}. To solve the overfitting problem, we adapt to calibrate the confidence score of the classifier (Inception-v3), to which we opt to apply the popular network calibration method of temperature scaling~\cite{guo2017calibration}. The formula of I-FID is defined below: 
\begin{equation}
    \mathrm{I-FID}=||\mu_r-\mu_g||^2+\mathrm{trace}\left(\Sigma_r+\Sigma_g-2(\Sigma_r\Sigma_g)^{\frac{1}{2}}\right)
\end{equation}
where $X_r\sim\mathcal{N}(\mu_r,\Sigma_r)$ and $X_g\sim\mathcal{N}(\mu_g,\Sigma_g)$ are the features of real images and generated images extracted by a pre-trained Inception-v3 model. Then, these two feature sets are represented as two multivariate Gaussian distributions. A lower I-FID value indicates superior image realism. The formula of I-IS is defined below: 
\begin{equation}
    \mathrm{I-IS}=\exp(\mathbb{E}_{x}D_{KL}(p(y_c|x)\parallel p(y_c))),
\end{equation}
where $x$ is the generated image and $y_c$ is the class label. The KL-divergence between $p(y_c)|x)$ and $p(y_c|x)$ should be large. Higher IS value means better image quality and diversity.

The RP metric is widely utilized to assess the consistency between text and image. The principle behind RP involves re-querying a synthesized image using the initial input caption. Specifically, an image is generated based on a true textual description amidst 99 other randomly chosen mismatched captions. This generated image is then used to search for the input description among 100 potential captions. The retrieval is considered successful if the image's matching score with the original caption ranks the highest. A higher RP value indicates a better degree of text-image matching.

Secondly, we define the error-detection ratio of entity ($Error_e$) and relationship ($Error_r$) based on the MR specified in Sec~\ref{MO}. 

\begin{equation}
    Error_e=\frac{\sum(p_e)}{N},
\end{equation}
\begin{equation}
    Error_r=\frac{\sum(p_r)}{N},
\end{equation}
where $p_e$ and $p_r$ represent the error reports for entities and relationships according to the certain MR. $N$ denotes the sum of the test samples. A higher error report rate indicates the detection of more defects. 

Thirdly, we apply the accuracy referring to object detection and relationship retrieval for both entities and relationships. We use the scene graph generation model~\cite{tang2020unbiased} (The model possesses a detection accuracy rate of 90\%) to detect the generated images. To more vividly demonstrate our testing method, we use 1 minus the accuracy rate to represent the miss-detection ratio. The two metrics are formulated as follows:
\begin{equation}
    Miss_e = 1-\frac{\sum{D_e(r)}}{\sum{Set_e(s)}}
\label{misse}
\end{equation}

\begin{equation}
    Miss_r = 1-\frac{\sum{D_r(r)}}{\sum{Set_r(s)}}
\label{missr}
\end{equation}
where $s$ and $r$ represent the input text and generated image. $Set_r(s)$ and $Set_e(s)$ donate the relationships and entities contained in the input text. $D_e(r)$ and $D_r(r)$ represent the detection results of entities and relationships by model $D_e$ and $D_r$. A higher miss-detection ratio means that more focal entities and relationships are lost in the generated images.

\section{Result Analysis}
\subsection{Answer to RQ1}

\begin{table}
\centering
\caption{The I-FID, I-IS and RP of different software on the tests generated by five guidance approaches and original data.}
\label{rq1}
\resizebox{0.6\linewidth}{!}{
\begin{tabular}{c|c|ccc} 
\toprule
\textbf{Software}     & \textbf{Oper} & \textbf{I-FID} & \textbf{I-IS} & \textbf{RP}        \\ 
\midrule
\multirow{5}{*}{Stable Diffusion XL}& Orig & 24.27 & 45.83 & 94.09\%  \\ 
                       & SS   & 25.63 & 44.09 & 92.07\%  \\
                      & ER\_R & 25.54 & 41.80 & 86.95\%  \\ 
                      & ER\_A & \textbf{25.67} & \textbf{38.65} & 87.05\%  \\ 
                      & EC   & 25.60 & 41.46 & \textbf{77.18\%} \\
\midrule
\multirow{5}{*}{DeepAI} & Orig & 26.41 & 44.38 & 93.01\%  \\ 
                       & SS   & 27.68 & 43.47 & 90.35\%  \\ 
                    & ER\_R & 28.36 & 43.62 & 86.39\%  \\ 
                & ER\_A & \textbf{28.60} & \textbf{40.49} & 86.66\%  \\ 
                      & EC   & 28.03 & 43.15 & \textbf{77.20\%}  \\
\midrule
\multirow{5}{*}{Wan Xiang} & Orig & 26.16 & 48.38 & 94.02\%  \\ 
                        & SS   & 28.53 & 48.78 & 92.79\%  \\ 
                    & ER\_R & 32.15 & 45.14 & 88.06\%  \\ 
                      & ER\_A & \textbf{33.11} & \textbf{41.94} & 86.24\%  \\ 
                      & EC   & 27.99 & 44.16 & \textbf{78.56\%}\\ 
\midrule
\multirow{5}{*}{OpenJourney}  & Orig & 30.36 & 43.80 & 91.68\%  \\ 
                        & SS   & 30.97 & 42.64 & 87.47\%  \\
                 & ER\_R & 31.29 & 42.52 & 82.99\%  \\ 
                      & ER\_A & \textbf{31.49} & \textbf{40.29} & 81.91\%  \\ 
                      & EC   & 30.76 & 41.14 & \textbf{72.33\%}  \\
\midrule
\multirow{5}{*}{DALLE} & Orig & 36.60 & 39.70 & 93.89\%  \\ 
                & SS   & 37.62 & 38.98 & 92.29\%  \\
               & ER\_R & 38.35 & 39.03 & 91.29\%  \\ 
              & ER\_A & \textbf{39.17} & \textbf{34.34} & 90.41\%  \\ 
              & EC   & 38.67 & 34.90 & \textbf{84.83\%}  \\ 
\midrule
MS COCO               & Real-image & 2.62 & 51.25 & 95.54\% \\
\bottomrule
\end{tabular}
}
\end{table}

We conduct our ACTesting on each software to explicitly demonstrate its effectiveness. we implement three mutation operators (EC, ER\_R, ER\_A) and one baseline mutation operator (SS) to generate four mutation testing sets. We also display the testing results of the metrics on real images for comparison~\cite{dinh2022tise}.
\textbf{Results.} Table~\ref{rq1} presents the I-FID, I-IS, and RP of all tested software on the seed sets and mutation testing sets. It can be seen that the three metrics of the baseline SS operator reflect a certain degree of decrease in generation image quality. The three mutation operators of ACTesting perform better. From the third column in Table~\ref{rq1}, it's evident that our operators effectively increased the I-FID values, thereby reducing the quality of the generated images. Among them, the ER\_A operator showed an average increase of about 5\%, ranking first in effectiveness. Additionally, the I-IS values for all the software under test also decrease. The decreased range of EC, ER\_A, and ER\_R operators is 2.9\% to 9.6\%, 8\% to 15\%, and 1.7\% to 8.8\%. In the RP column, the three mutation operators significantly reduced the precision values of the seed test set, with the EC operator consistently decreasing the RP value by 16.4\% to 21.1\%. The ER\_R and ER\_A operators, on average, decreased by 7.5\% and 8.4\% respectively.

\textbf{Analysis.} From the results, the three operators included in ACTesting all outperform the baseline SS. The test sets mutated by the ER\_A operator achieve relatively higher I-FID values and lower I-IS values. This is because the ER\_A operator introduces unfamiliar triplets and increases the difficulty and complexity of the generation process. Additionally, the test set mutated by the EC operator leads to a consistent and substantial decrease in RP values. This is attributed to the EC operator disrupting the natural ER triples, compelling the software to generate more innovative results. Consequently, this reduces the consistency between the text and the generated images. The reason for the lesser decline in the I-FID and I-IS metrics compared to RP is that our operators are primarily focused on detecting errors in text-image consistency. The results indicate that ACTesting can generate tests that reduce the quality of T2I software output.

\subsection{Answer to RQ2}
\begin{figure}[!tbp]
  \centering
  \includegraphics[width=0.75\linewidth]{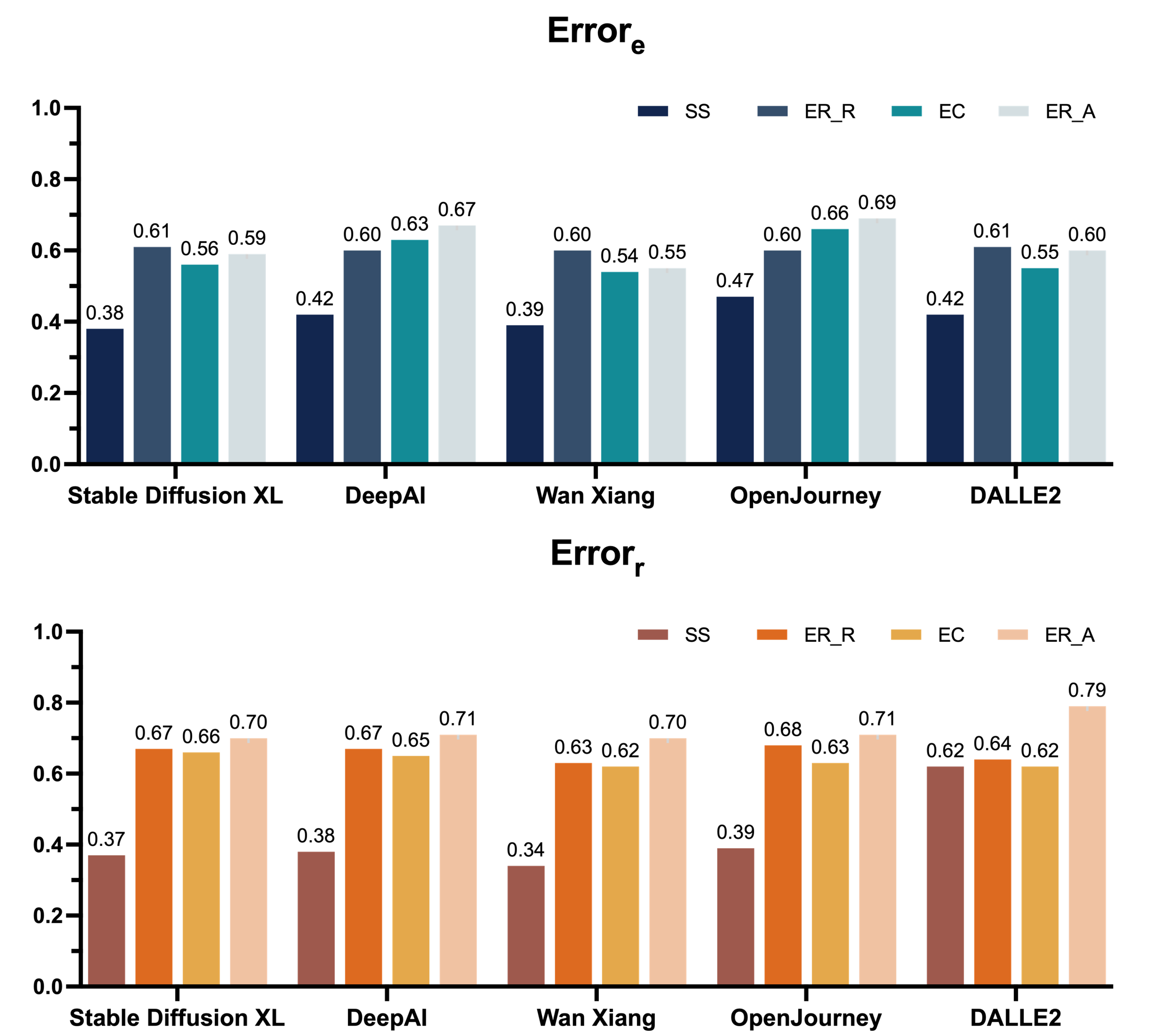}
  \caption{The error rates of different software on the tests generated by four mutation operators.}
  \label{rq2}
\end{figure}
As described in Sec~\ref{MO}, we treat MR as the practical adequacy criteria, including definitions for entities and relationships. Therefore, the testing results are separately addressed in error reporting, calculated by the error rate of entities and relationships ($Error_e$ and $Error_r$). We define error detection of the SS operator when entities or relationships of the output are not exactly consistent with the input.

\textbf{Results.} Fig.~\ref{rq2} presents the detection error rates of five T2I software on the test inputs generated by four operators (three proposed operators and one baseline). The upper part shows the error rate of entities. The error rate for the SS operator remains around 0.40, while the ER\_R and EC operators consistently stay around 0.60. The ER\_A operator often ranks the highest, reaching its highest error rate of 0.69 in the OpenJourney software. The lower part illustrates the error rates for MR of relationships. In this case, the SS operator maintains error rates all below 0.4, while both the ER\_R and EC operators exhibit error rates all above 0.6 but below 0.7. The ER\_A operator ranks highest, maintaining an error rate of 0.7. 

\textbf{Analysis.}
Although the SS operator can also detect errors within a certain range, our proposed operators can detect more erroneous behaviors of the tested software. The testing efficiency of the ER\_A operator is around 1.75 that of the SS operator. The reason is that ACTesting transforms entities and relationships, which allows the new text to encompass new testing scenarios. From the experimental results of $Error_e$, ER\_A is effective for most software on entities and relationships. The results of $Error_r$ demonstrate that both of the three operators can detect the errors effectively and stably. In addition, By comparing with the results in Table~\ref{rq1}, it is evident that the differences from the baseline reflected by the $Error_e$ and $Error_r$ are more pronounced. This is because the two metrics more accurately indicate the errors present in the generated images (the specific types of errors are described in Sec~\ref{Errors}).

\begin{table}
\centering
\caption{The ablation experiment results of different mutation operators. Bold represents the highest missed detection rate, and underline represents the second-ranking.}
\label{rq3}
\resizebox{\linewidth}{!}{
\begin{tabular}{c|cccccc|cccccc} 
\toprule
                    & \multicolumn{6}{c|}{$Miss_e$}                                              & \multicolumn{6}{c}{$Miss_r$}                                               \\ 
\midrule
Software            & Orig   & ER\_R   & ER\_A   & EC     & EC+ER\_R        & EC+ER\_A         & Orig   & ER\_R   & ER\_A   & EC     & EC+ER\_R        & EC+ER\_A          \\ 
SD XL & 0.1834 & 0.3122 & 0.3325 & 0.3528 & \underline{0.3702} & \textbf{0.4338} & 0.3227 & 0.2986 & 0.3332 & 0.3329 & \underline{0.3443} & \textbf{0.3528}  \\ 
DeepAI              & 0.2076 & 0.3086 & 0.3732 & 0.3675 & \underline{0.3904} & \textbf{0.4755} & 0.3206 & 0.3107 & 0.3261 & 0.3430 & \underline{0.3435} & \textbf{0.3580}  \\ 
Wan Xiang           & 0.1844 & 0.2733 & 0.3271 & 0.3449 & \underline{0.3598} & \textbf{0.4293} & 0.3360 & 0.3164 & 0.3334 & 0.3433 & \underline{0.3506} & \textbf{0.3541}  \\ 
OpenJourney         & 0.2364 & 0.2876 & 0.4099 & 0.4011 & \underline{0.4273} & \textbf{0.5210} & 0.3395 & 0.3301 & 0.3463 & 0.3464 & \underline{0.3520} & \textbf{0.3699}  \\
DALLE         & 0.2060 & 0.1929 & 0.3553 & 0.3666 & \underline{0.3929} & \textbf{0.4497} & 0.3420 & 0.3329 & 0.3511 & 0.3647 & \underline{0.3893} & \textbf{0.3940}  \\
\bottomrule
\end{tabular}
}
\end{table}

\subsection{Answer to RQ3}
To better demonstrate the effectiveness of the three kinds of operators, we conduct ablation experiments. The evaluation criteria are the miss-detection rate for entities and relationships in the generated images. Due to the conflict between ER\_R and ER\_A operators, we combine the EC operator with the ER\_R operator and ER\_A operator, respectively. Table~\ref{rq3} shows the ablation experiment results of five T2I software.

\textbf{Results.} The $Miss_e$ column displays the miss-detection rate of entities under various mutation conditions. The table clearly shows that the tests conducted with the combined EC plus ER\_R, and EC plus ER\_A operators result in the highest missed detection rates, surpassing those of the three individual operators. The value of $Miss_e$ for the EC+ER\_A operator is, on average, 26\% higher than that of the EC operator and 29.1\% higher than that of the ER\_A operator. The average value of $Miss_e$ for the EC+ER\_R operator exceeds that of the EC operator by 5.5\% and surpasses the ER\_R operator by a significant margin of 31.3\%. The $Miss_r$ column reflects the rate of missed detections for relationships. It shows a trend consistent with that of the $Miss_e$ column. Specifically, the average $Miss_r$ value for the EC+ER\_A operator exceeds those of the EC and ER\_A operators by 5.1\% and 7.2\%, respectively. Although the EC+ER\_R operator leads the EC operator on average $Miss_r$ by only 1.8\%, it surpasses the ER\_R operator by a notable margin of 10.8\%, still demonstrating significant competitiveness.

\textbf{Analysis.} 
The ablation study demonstrates that three kinds of operators are effective and flexible to use. Firstly, the value of $Miss_e$ and $Miss_r$ for each mutation operator is substantially higher than that of the seed test set, where EC+ER\_A shows the best performance. This aligns with the trends observed in the common metrics presented in Table~\ref{rq1}, further validating the effectiveness of this testing method. Secondly, we observe that the missed detection rate for entity is higher than that for relationship. This may be attributed to the higher precision of the object detection model and indicates that the accuracy of relationship recognition in image scene understanding still requires improvement.

\section{Discussion}

\subsection{False Positive and Negative}
\textbf{ACTesting suffers little from false positives and negatives.} Due to the large number of experimental results, we randomly select 100 images for manual review. Although ACTesting utilizes high-precision detection and recognition models to calculate and analyze the testing results, we still encounter low false positive and false negative rates. Due to the strict setting of the MR, most of the false positive errors depend on the accuracy of the detection model; that is, the recognition model may mistake the background for the focal entity (the proportion is lower than 10\%). The majority of false negatives of ACTesting come from the recognition model's perception that the entity is partly distorted (Image~4 in Fig.~\ref{motivation}), making it unable to be identified as a focal entity. Therefore, the false negative depends on the threshold of the detection models (8\% in our setting).

\subsection{Types of Detected Errors}
\label{Errors}
\begin{figure}[!tbp]
  \centering
  \begin{subfigure}[b]{0.95\linewidth}
    \includegraphics[width=\linewidth]{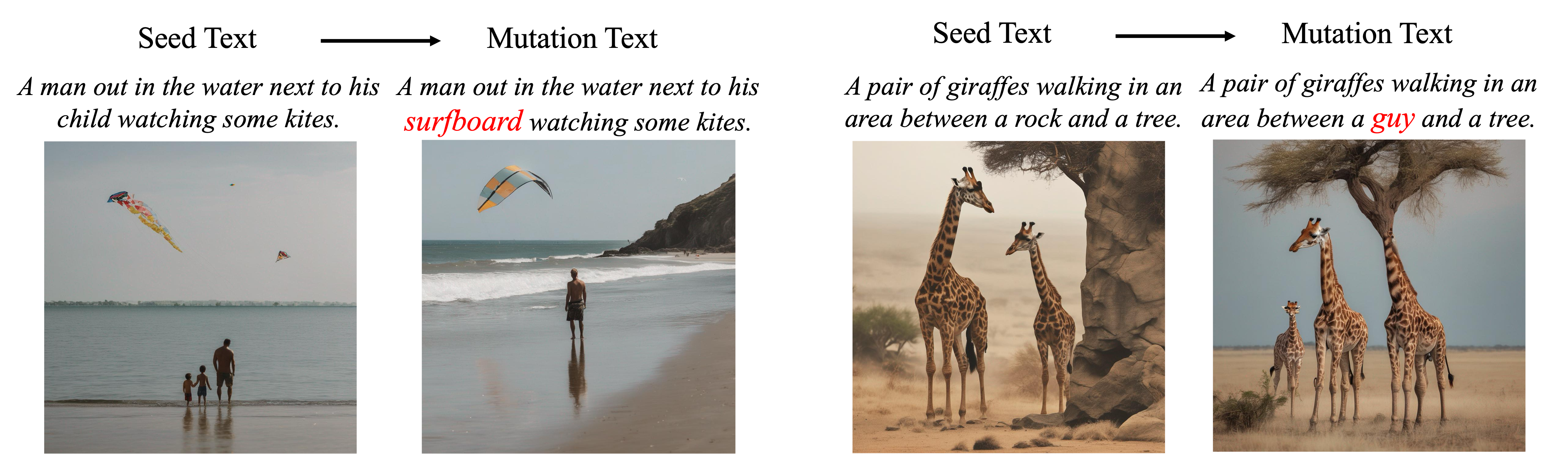}
    \caption{Entity Omission Error. The entities highlighted in red are absent in the images.}
    \label{fig:cs1}
  \end{subfigure}
  
  \begin{subfigure}[b]{0.95\linewidth}
    \includegraphics[width=\linewidth]{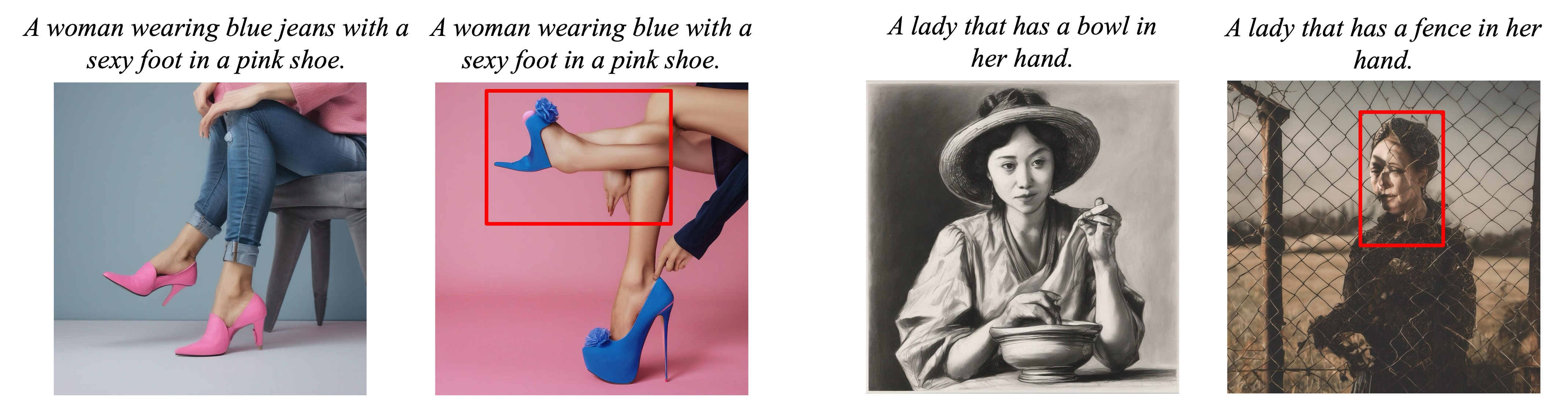}
    \caption{Entity Distortion Error. The distortion part is outlined in red in the images)}
    \label{fig:cs2}
  \end{subfigure}

  \begin{subfigure}[b]{0.95\linewidth}
    \includegraphics[width=\linewidth]{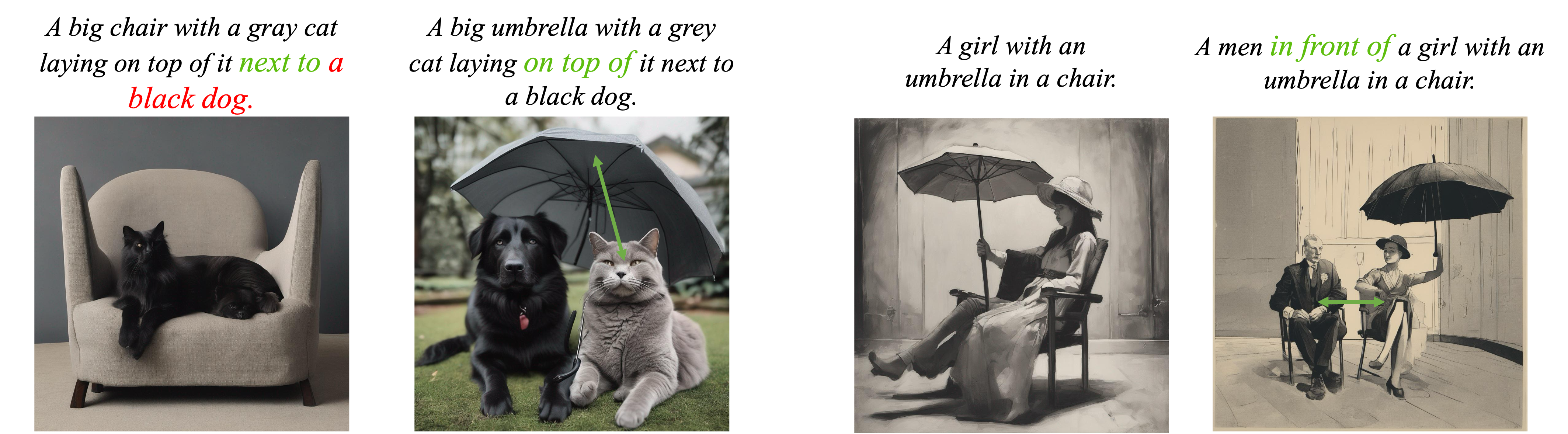}
    \caption{Relationship Construction Error. The relationships marked in green are incorrectly represented in the images.}
    \label{fig:cs3}
  \end{subfigure}
  
  \caption{Examples of each type of error reported by ACTesting. Images are generated by Stable Diffusion.}
  \label{fig:cs}
\end{figure}

To further illustrate the errors detected by ACTesting, We provide two examples for each of the three error types, along with an indication of the reasons for the errors in Fig.~\ref{fig:cs}. Firstly, the most prevalent error type is \textbf{Entity Omission Error}, commonly occurring when ACTesting employs the EC operator. As illustrated in Fig.~\ref{fig:cs}~(a), the entity ``child'' is replaced with ``surfboard'' and ``rock'' is replaced with ``guy'' in the text; however, the generated images omit these entities. Secondly, while the T2I software is capable of generating entities, the quality of the generation is unsatisfactory. Fig.~\ref{fig:cs}~(b) presents examples of \textbf{Entity Distortion Error}, which can be tested by all mutation operators. These two types of errors are related to entities. Both missing and distorted image entities are classified as entity generation failures by the object detection model, contributing to $Error_e$ and $Miss_e$. The third error type is \textbf{Relationship Construction Error}, which indicates incorrect relationship representation between entities in the images (e.g., the grey cat is ``under'' the umbrella rather than ``on top of'' it in Fig.~\ref{fig:cs}~(c)), contributing to $Error_r$ and $Miss_r$. It is notable that (1) the three types of error can occur at the same time, as shown in the first image in Fig.~\ref{fig:cs}~(c); (2) If an entity is classified as a generation failure, then any relationships associated with that entity will also be deemed failures.

\subsection{Analysis and Insight}
\textbf{ER relationships can be incorporated into the construction and training of generative models.} When we abstract the text-image pairs into ER triples, for text, it is not necessary to describe all relationships between entities, while for images, it is necessary to present all of them when constructing a visual scene. This difference in presentation formats can lead generative software to interpret the text as triples in different paradigms, leading to omission during the generation process. Complex or unreal-world ERs in the input text can pose significant challenges for T2I model training on natural scenes. Regarding the most commonly used diffusion models, which are based on modeling noise probability distributions, their loss function constrains the predicted noise to be close to the true noise at each step. When there are more or unfamiliar elements and the joint probability distribution becomes more complex, the denoising process becomes more challenging. Therefore, if ER relationships can be used as guided input during the model construction phase, it may enhance the model's ability to construct the core parts of an image.

\section{Threats to Validity}
\textit{\textbf{Test Subject.}} The selection of T2I software is one of the vital threats to validity. With the rapid development of T2I software, a diverse array of such software is continually emerging, each with varying engines and mechanisms. To alleviate this threat, we employ five commonly used T2I software powered by different core engines. \\%
\textit{\textbf{Test Data Selection.}}
The selection of the source dataset is another threat. Because not all training and validation datasets used by various software are publicly available, there is a potential data domain unfairness. In addition, prompt construction based on the input text is crucial. To alleviate this threat, we use the most commonly used and classic open-source dataset, MS-COCO, as our seed collection to ensure a level of fairness in the experiments. We also do not include any additional description style or scene details in the system prompt to minimize the impact on the testing process.\\%
\textit{\textbf{Evaluation Model Selection.}}
The last threat is related to the deep learning models used during the evaluation process. To overcome potential biases, we use the regularized Inception-v3 model following~\cite{dinh2022tise} to avoid the overfitting problem and apply the open-source pre-trained weights of RvlBert and CLIP.

\section{Related Work}
This section covers the evaluation method of T2I software and discusses various AI testing methods relevant to our study.

\subsection{Text-to-Image Evaluation}

Researchers propose a series of evaluation metrics and methods to reasonably assess the performance of T2I software, which are usually based on image quality and text-image alignment. Inception score (IS)~\cite{salimans2016improved} and Fréchet Inception Distance (FIS)~\cite{heusel2017gans} are two common indicators to leverage the image quality and image diversity based on pre-trained Inception-v3 network~\cite{szegedy2016rethinking}. R-precision (RP)~\cite{xu2018attngan} is usually used to assess the consistency between text and images. Dinh et al.~\cite{dinh2022tise} propose several improved metrics based on these metrics. Cho et al.~\cite{cho2023visual} introduce two novel interpretable visual programming frameworks for T2I generation. DALL-Eval ~\cite{cho2023dall} measures three visual reasoning skills, including object recognition and counting, and spatial relation understanding to evaluate T2I models. Saxon et al.~\cite{saxon2024evaluates} introduce T2IScoreScore, a dataset for rigorously evaluating t2i prompt faithfulness metrics. Wan et al. ~\cite{wan2024factuality} propose DoFaiR, a benchmark to measure the balance between diversity interventions and demographic accuracy in Text-to-Image models. FlashEval~\cite{zhao2024flasheval} is proposed recently to increase the evaluation efficiency by selecting the representative subset of the text-image dataset.

\subsection{AI Testing}
Due to the scarcity of testing methods for the T2I task, we introduce some relevant testing methods. After achieving success in traditional testing tasks with metamorphic testing, researchers propose testing methods to test the deep learning systems. Xie et al. ~\cite{xie2020mettle} develop a metamorphic testing approach to assessing and validating unsupervised machine learning systems. Ma et al.~\cite{Ma2018} apply the metamorphic testing method and specialize it for deep learning systems to assess the quality of test data. Berend et al.~\cite{Berend2020} conduct a rich empirical study identifying the impact of mutation operators and coverage criteria on the distribution of the generated deep learning test cases. Chang et al. ~\cite{chang2024veglue} introduce VEglue, an object-aligned joint erasing approach for testing Visual Entailment (VE) systems. AI testing methods cover various areas, including autonomous driving~\cite{Jana2018, Feng2021, Li2022, gao2023benchmarking}, speech recognition systems~\cite{Ji2022}, question-answering systems~\cite{Liu2022, xie2023qaasker+}, machine translation~\cite{zhang2024machine, he2021testing,sun2020automatic}, and image captioning systems~\cite{Yu2022, xie2024metamorphic}.

However, aside from the issue of overfitting inherent in evaluation metrics themselves, T2I software lacks systematic testing methods to assess its performance. Unlike the previous methods, ACTesting aims to apply the metamorphic testing theory and adaptability density constraint, utilizing the ER triple to present the cross-modal semantic information. It constructs the mutation input text and treats the MR as detection criteria to reveal hidden errors of tested T2I software.

\section{Conclusion}
In this paper, we introduce ACTesting, an automated, black-box approach grounded in metamorphic testing theory to address the challenges in cross-modal testing for Text-to-Image (T2I) software. ACTesting emphasizes maintaining semantic consistency across different modalities and utilizes the ER triple to encapsulate key semantic information. We develop three mutation operators based on metamorphic testing theory and adaptability density constraint to increase test scenarios. Our experiments involve five T2I software, generating 142,170 synthetic images. The findings reveal that ACTesting could degrade image quality by 2.9\% to 15\% and diminish text-image match consistency by 7.5\% to 21.1\%. ACTesting identifies the deficiency of the current T2I software in dealing with unrealistic and complex scenarios. In the future, the quality of the T2I software can be improved by retraining it using input text based on our mutation operators. We also plan to extend ACTesting to a broader range of creative generative artificial intelligence software.

\section{Acknowledgement}
This work is supported partially by the National Natural Science Foundation of China (61932012, 62372228).

\bibliographystyle{ACM-Reference-Format}

\begin{thebibliography}{64}

  
  \ifx \showCODEN    \undefined \def \showCODEN     #1{\unskip}     \fi
  \ifx \showDOI      \undefined \def \showDOI       #1{#1}\fi
  \ifx \showISBNx    \undefined \def \showISBNx     #1{\unskip}     \fi
  \ifx \showISBNxiii \undefined \def \showISBNxiii  #1{\unskip}     \fi
  \ifx \showISSN     \undefined \def \showISSN      #1{\unskip}     \fi
  \ifx \showLCCN     \undefined \def \showLCCN      #1{\unskip}     \fi
  \ifx \shownote     \undefined \def \shownote      #1{#1}          \fi
  \ifx \showarticletitle \undefined \def \showarticletitle #1{#1}   \fi
  \ifx \showURL      \undefined \def \showURL       {\relax}        \fi
  \providecommand\bibfield[2]{#2}
  \providecommand\bibinfo[2]{#2}
  \providecommand\natexlab[1]{#1}
  \providecommand\showeprint[2][]{arXiv:#2}
  
  \bibitem[Alibaba(2023)]%
          {WanXiang}
  \bibfield{author}{\bibinfo{person}{Alibaba}.} \bibinfo{year}{2023}\natexlab{}.
  \newblock \bibinfo{booktitle}{\emph{Wan Xiang}}.
  \newblock
  \urldef\tempurl%
  \url{https://tongyi.aliyun.com/wanxiang/}
  \showURL{%
  \tempurl}
  
  
  \bibitem[Baidu(2023)]%
          {baidu}
  \bibfield{author}{\bibinfo{person}{Baidu}.} \bibinfo{year}{2023}\natexlab{}.
  \newblock \bibinfo{title}{ERNIE-ViLG}.
  \newblock \bibinfo{howpublished}{Website}.
  \newblock
  \newblock
  \shownote{\url{https://ai.baidu.com/tech/creativity/ernie_Vilg}}.
  
  
  \bibitem[Berend et~al\mbox{.}(2020)]%
          {Berend2020}
  \bibfield{author}{\bibinfo{person}{David Berend}, \bibinfo{person}{Xiaofei Xie}, \bibinfo{person}{Lei Ma}, \bibinfo{person}{Lingjun Zhou}, \bibinfo{person}{Yang Liu}, \bibinfo{person}{Chi Xu}, {and} \bibinfo{person}{Jianjun Zhao}.} \bibinfo{year}{2020}\natexlab{}.
  \newblock \showarticletitle{Cats are not fish: Deep learning testing calls for out-of-distribution awareness}.
  \newblock  (\bibinfo{year}{2020}), \bibinfo{pages}{1041--1052}.
  \newblock
  
  
  \bibitem[Borji(2019)]%
          {BORJI201941}
  \bibfield{author}{\bibinfo{person}{Ali Borji}.} \bibinfo{year}{2019}\natexlab{}.
  \newblock \showarticletitle{Pros and cons of GAN evaluation measures}.
  \newblock \bibinfo{journal}{\emph{Computer Vision and Image Understanding}}  \bibinfo{volume}{179} (\bibinfo{year}{2019}), \bibinfo{pages}{41--65}.
  \newblock
  \showISSN{1077-3142}
  \urldef\tempurl%
  \url{https://doi.org/10.1016/j.cviu.2018.10.009}
  \showDOI{\tempurl}
  
  
  \bibitem[Bynagari(2019)]%
          {heusel2017gans}
  \bibfield{author}{\bibinfo{person}{Naresh~Babu Bynagari}.} \bibinfo{year}{2019}\natexlab{}.
  \newblock \showarticletitle{GANs trained by a two time-scale update rule converge to a local Nash equilibrium}.
  \newblock \bibinfo{journal}{\emph{Asian Journal of Applied Science and Engineering}}  \bibinfo{volume}{8} (\bibinfo{year}{2019}), \bibinfo{pages}{25--34}.
  \newblock
  
  
  \bibitem[Caesar et~al\mbox{.}(2018)]%
          {caesar2018coco}
  \bibfield{author}{\bibinfo{person}{Holger Caesar}, \bibinfo{person}{Jasper Uijlings}, {and} \bibinfo{person}{Vittorio Ferrari}.} \bibinfo{year}{2018}\natexlab{}.
  \newblock \showarticletitle{Coco-stuff: Thing and stuff classes in context}. In \bibinfo{booktitle}{\emph{Proceedings of the IEEE conference on computer vision and pattern recognition}}. \bibinfo{pages}{1209--1218}.
  \newblock
  
  
  \bibitem[Chang et~al\mbox{.}(2024)]%
          {chang2024veglue}
  \bibfield{author}{\bibinfo{person}{Zhiyuan Chang}, \bibinfo{person}{Mingyang Li}, \bibinfo{person}{Junjie Wang}, \bibinfo{person}{Cheng Li}, {and} \bibinfo{person}{Qing Wang}.} \bibinfo{year}{2024}\natexlab{}.
  \newblock \showarticletitle{VEglue: Testing Visual Entailment Systems via Object-Aligned Joint Erasing}.
  \newblock \bibinfo{journal}{\emph{CoRR}}  \bibinfo{volume}{abs/2403.02581} (\bibinfo{year}{2024}).
  \newblock
  
  
  \bibitem[Chiou et~al\mbox{.}(2021)]%
          {rvlbert}
  \bibfield{author}{\bibinfo{person}{Meng-Jiun Chiou}, \bibinfo{person}{Roger Zimmermann}, {and} \bibinfo{person}{Jiashi Feng}.} \bibinfo{year}{2021}\natexlab{}.
  \newblock \showarticletitle{Visual Relationship Detection With Visual-Linguistic Knowledge From Multimodal Representations}.
  \newblock \bibinfo{journal}{\emph{IEEE Access}}  \bibinfo{volume}{9} (\bibinfo{year}{2021}), \bibinfo{pages}{50441--50451}.
  \newblock
  
  
  \bibitem[Cho et~al\mbox{.}(2023a)]%
          {cho2023dall}
  \bibfield{author}{\bibinfo{person}{Jaemin Cho}, \bibinfo{person}{Abhay Zala}, {and} \bibinfo{person}{Mohit Bansal}.} \bibinfo{year}{2023}\natexlab{a}.
  \newblock \showarticletitle{Dall-eval: Probing the reasoning skills and social biases of text-to-image generation models}. In \bibinfo{booktitle}{\emph{Proceedings of the IEEE/CVF International Conference on Computer Vision}}. \bibinfo{pages}{3043--3054}.
  \newblock
  
  
  \bibitem[Cho et~al\mbox{.}(2023b)]%
          {cho2023visual}
  \bibfield{author}{\bibinfo{person}{Jaemin Cho}, \bibinfo{person}{Abhay Zala}, {and} \bibinfo{person}{Mohit Bansal}.} \bibinfo{year}{2023}\natexlab{b}.
  \newblock \showarticletitle{Visual Programming for Step-by-Step Text-to-Image Generation and Evaluation}. In \bibinfo{booktitle}{\emph{NeurIPS}}.
  \newblock
  
  
  \bibitem[DeepAI(2023)]%
          {DeepAI}
  \bibfield{author}{\bibinfo{person}{DeepAI}.} \bibinfo{year}{2023}\natexlab{}.
  \newblock \bibinfo{title}{DeepAI Image Generator}.
  \newblock \bibinfo{howpublished}{Website}.
  \newblock
  \newblock
  \shownote{\url{https://deepai.org/machine-learning-model/text2img}}.
  
  
  \bibitem[Devlin et~al\mbox{.}(2019)]%
          {devlin2018bert}
  \bibfield{author}{\bibinfo{person}{Jacob Devlin}, \bibinfo{person}{Ming{-}Wei Chang}, \bibinfo{person}{Kenton Lee}, {and} \bibinfo{person}{Kristina Toutanova}.} \bibinfo{year}{2019}\natexlab{}.
  \newblock \showarticletitle{{BERT:} Pre-training of Deep Bidirectional Transformers for Language Understanding}. In \bibinfo{booktitle}{\emph{{NAACL-HLT} {(1)}}}. \bibinfo{publisher}{Association for Computational Linguistics}, \bibinfo{pages}{4171--4186}.
  \newblock
  
  
  \bibitem[Dinh et~al\mbox{.}(2022)]%
          {dinh2022tise}
  \bibfield{author}{\bibinfo{person}{Tan~M Dinh}, \bibinfo{person}{Rang Nguyen}, {and} \bibinfo{person}{Binh-Son Hua}.} \bibinfo{year}{2022}\natexlab{}.
  \newblock \showarticletitle{TISE: Bag of metrics for text-to-image synthesis evaluation}. In \bibinfo{booktitle}{\emph{European Conference on Computer Vision}}. \bibinfo{pages}{594--609}.
  \newblock
  
  
  \bibitem[Feng et~al\mbox{.}(2020)]%
          {Feng2021}
  \bibfield{author}{\bibinfo{person}{Di Feng}, \bibinfo{person}{Christian Haase-Sch{\"u}tz}, \bibinfo{person}{Lars Rosenbaum}, \bibinfo{person}{Heinz Hertlein}, \bibinfo{person}{Claudius Glaeser}, \bibinfo{person}{Fabian Timm}, \bibinfo{person}{Werner Wiesbeck}, {and} \bibinfo{person}{Klaus Dietmayer}.} \bibinfo{year}{2020}\natexlab{}.
  \newblock \showarticletitle{Deep multi-modal object detection and semantic segmentation for autonomous driving: Datasets, methods, and challenges}.
  \newblock \bibinfo{journal}{\emph{IEEE Transactions on Intelligent Transportation Systems}} \bibinfo{volume}{22}, \bibinfo{number}{3} (\bibinfo{year}{2020}), \bibinfo{pages}{1341--1360}.
  \newblock
  
  
  \bibitem[Frolov et~al\mbox{.}(2021)]%
          {frolov2021adversarial}
  \bibfield{author}{\bibinfo{person}{Stanislav Frolov}, \bibinfo{person}{Tobias Hinz}, \bibinfo{person}{Federico Raue}, \bibinfo{person}{J{\"o}rn Hees}, {and} \bibinfo{person}{Andreas Dengel}.} \bibinfo{year}{2021}\natexlab{}.
  \newblock \showarticletitle{Adversarial text-to-image synthesis: A review}.
  \newblock \bibinfo{journal}{\emph{Neural Networks}}  \bibinfo{volume}{144} (\bibinfo{year}{2021}), \bibinfo{pages}{187--209}.
  \newblock
  
  
  \bibitem[Gao et~al\mbox{.}(2023)]%
          {gao2023benchmarking}
  \bibfield{author}{\bibinfo{person}{Xinyu Gao}, \bibinfo{person}{Zhijie Wang}, \bibinfo{person}{Yang Feng}, \bibinfo{person}{Lei Ma}, \bibinfo{person}{Zhenyu Chen}, {and} \bibinfo{person}{Baowen Xu}.} \bibinfo{year}{2023}\natexlab{}.
  \newblock \showarticletitle{Benchmarking Robustness of AI-Enabled Multi-sensor Fusion Systems: Challenges and Opportunities}. In \bibinfo{booktitle}{\emph{{ESEC/SIGSOFT} {FSE}}}. \bibinfo{publisher}{{ACM}}, \bibinfo{pages}{871--882}.
  \newblock
  
  
  \bibitem[Gu et~al\mbox{.}(2022)]%
          {Gu2022}
  \bibfield{author}{\bibinfo{person}{Shuyang Gu}, \bibinfo{person}{Dong Chen}, \bibinfo{person}{Jianmin Bao}, \bibinfo{person}{Fang Wen}, \bibinfo{person}{Bo Zhang}, \bibinfo{person}{Dongdong Chen}, \bibinfo{person}{Lu Yuan}, {and} \bibinfo{person}{Baining Guo}.} \bibinfo{year}{2022}\natexlab{}.
  \newblock \showarticletitle{Vector quantized diffusion model for text-to-image synthesis}. In \bibinfo{booktitle}{\emph{Proceedings of the IEEE/CVF Conference on Computer Vision and Pattern Recognition}}. \bibinfo{pages}{10696--10706}.
  \newblock
  
  
  \bibitem[Guo et~al\mbox{.}(2017)]%
          {guo2017calibration}
  \bibfield{author}{\bibinfo{person}{Chuan Guo}, \bibinfo{person}{Geoff Pleiss}, \bibinfo{person}{Yu Sun}, {and} \bibinfo{person}{Kilian~Q Weinberger}.} \bibinfo{year}{2017}\natexlab{}.
  \newblock \showarticletitle{On calibration of modern neural networks}. In \bibinfo{booktitle}{\emph{International conference on machine learning}}. \bibinfo{pages}{1321--1330}.
  \newblock
  
  
  \bibitem[He et~al\mbox{.}(2021)]%
          {he2021testing}
  \bibfield{author}{\bibinfo{person}{Pinjia He}, \bibinfo{person}{Clara Meister}, {and} \bibinfo{person}{Zhendong Su}.} \bibinfo{year}{2021}\natexlab{}.
  \newblock \showarticletitle{Testing machine translation via referential transparency}. In \bibinfo{booktitle}{\emph{2021 IEEE/ACM 43rd International Conference on Software Engineering (ICSE)}}. IEEE, \bibinfo{pages}{410--422}.
  \newblock
  
  
  \bibitem[Hinz et~al\mbox{.}(2020)]%
          {hinz2020semantic}
  \bibfield{author}{\bibinfo{person}{Tobias Hinz}, \bibinfo{person}{Stefan Heinrich}, {and} \bibinfo{person}{Stefan Wermter}.} \bibinfo{year}{2020}\natexlab{}.
  \newblock \showarticletitle{Semantic object accuracy for generative text-to-image synthesis}.
  \newblock \bibinfo{journal}{\emph{IEEE transactions on pattern analysis and machine intelligence}} \bibinfo{volume}{44}, \bibinfo{number}{3} (\bibinfo{year}{2020}), \bibinfo{pages}{1552--1565}.
  \newblock
  
  
  \bibitem[Huang et~al\mbox{.}(2023)]%
          {huang2023composer}
  \bibfield{author}{\bibinfo{person}{Lianghua Huang}, \bibinfo{person}{Di Chen}, \bibinfo{person}{Yu Liu}, \bibinfo{person}{Yujun Shen}, \bibinfo{person}{Deli Zhao}, {and} \bibinfo{person}{Jingren Zhou}.} \bibinfo{year}{2023}\natexlab{}.
  \newblock \showarticletitle{Composer: Creative and Controllable Image Synthesis with Composable Conditions}. In \bibinfo{booktitle}{\emph{{ICML}}} \emph{(\bibinfo{series}{Proceedings of Machine Learning Research}, Vol.~\bibinfo{volume}{202})}. \bibinfo{publisher}{{PMLR}}, \bibinfo{pages}{13753--13773}.
  \newblock
  
  
  \bibitem[Huang et~al\mbox{.}(2019)]%
          {huang2019realistic}
  \bibfield{author}{\bibinfo{person}{Wanming Huang}, \bibinfo{person}{Richard~Yi Da~Xu}, {and} \bibinfo{person}{Ian Oppermann}.} \bibinfo{year}{2019}\natexlab{}.
  \newblock \showarticletitle{Realistic image generation using region-phrase attention}. In \bibinfo{booktitle}{\emph{Asian Conference on Machine Learning}}. \bibinfo{pages}{284--299}.
  \newblock
  
  
  \bibitem[Huynh et~al\mbox{.}(2022)]%
          {medicial}
  \bibfield{author}{\bibinfo{person}{Tuan-Luc Huynh}, \bibinfo{person}{Khoi-Nguyen Nguyen-Ngoc}, \bibinfo{person}{Chi-Bien Chu}, \bibinfo{person}{Minh-Triet Tran}, {and} \bibinfo{person}{Trung-Nghia Le}.} \bibinfo{year}{2022}\natexlab{}.
  \newblock \showarticletitle{Multilingual Communication System with Deaf Individuals Utilizing Natural and Visual Languages}. In \bibinfo{booktitle}{\emph{2022 RIVF International Conference on Computing and Communication Technologies (RIVF)}}. \bibinfo{pages}{683--688}.
  \newblock
  \urldef\tempurl%
  \url{https://doi.org/10.1109/RIVF55975.2022.10013851}
  \showDOI{\tempurl}
  
  
  \bibitem[Ji et~al\mbox{.}(2022)]%
          {Ji2022}
  \bibfield{author}{\bibinfo{person}{Pin Ji}, \bibinfo{person}{Yang Feng}, \bibinfo{person}{Jia Liu}, \bibinfo{person}{Zhihong Zhao}, {and} \bibinfo{person}{Zhenyu Chen}.} \bibinfo{year}{2022}\natexlab{}.
  \newblock \showarticletitle{ASRTest: automated testing for deep-neural-network-driven speech recognition systems}.
  \newblock  (\bibinfo{year}{2022}), \bibinfo{pages}{189--201}.
  \newblock
  
  
  \bibitem[Kim et~al\mbox{.}(2022)]%
          {kim2022verse}
  \bibfield{author}{\bibinfo{person}{Taehoon Kim}, \bibinfo{person}{Gwangmo Song}, \bibinfo{person}{Sihaeng Lee}, \bibinfo{person}{Sangyun Kim}, \bibinfo{person}{Yewon Seo}, \bibinfo{person}{Soonyoung Lee}, \bibinfo{person}{Seung~Hwan Kim}, \bibinfo{person}{Honglak Lee}, {and} \bibinfo{person}{Kyunghoon Bae}.} \bibinfo{year}{2022}\natexlab{}.
  \newblock \showarticletitle{L-verse: Bidirectional generation between image and text}. In \bibinfo{booktitle}{\emph{Proceedings of the IEEE/CVF Conference on Computer Vision and Pattern Recognition}}. \bibinfo{pages}{16526--16536}.
  \newblock
  
  
  \bibitem[Lao et~al\mbox{.}(2019)]%
          {lao2019dual}
  \bibfield{author}{\bibinfo{person}{Qicheng Lao}, \bibinfo{person}{Mohammad Havaei}, \bibinfo{person}{Ahmad Pesaranghader}, \bibinfo{person}{Francis Dutil}, \bibinfo{person}{Lisa~Di Jorio}, {and} \bibinfo{person}{Thomas Fevens}.} \bibinfo{year}{2019}\natexlab{}.
  \newblock \showarticletitle{Dual adversarial inference for text-to-image synthesis}. In \bibinfo{booktitle}{\emph{Proceedings of the IEEE/CVF International Conference on Computer Vision}}. \bibinfo{pages}{7567--7576}.
  \newblock
  
  
  \bibitem[Lee et~al\mbox{.}(2024)]%
          {lee2024holistic}
  \bibfield{author}{\bibinfo{person}{Tony Lee}, \bibinfo{person}{Michihiro Yasunaga}, \bibinfo{person}{Chenlin Meng}, \bibinfo{person}{Yifan Mai}, \bibinfo{person}{Joon~Sung Park}, \bibinfo{person}{Agrim Gupta}, \bibinfo{person}{Yunzhi Zhang}, \bibinfo{person}{Deepak Narayanan}, \bibinfo{person}{Hannah Teufel}, \bibinfo{person}{Marco Bellagente}, {et~al\mbox{.}}} \bibinfo{year}{2024}\natexlab{}.
  \newblock \showarticletitle{Holistic evaluation of text-to-image models}.
  \newblock \bibinfo{journal}{\emph{Advances in Neural Information Processing Systems}}  \bibinfo{volume}{36} (\bibinfo{year}{2024}).
  \newblock
  
  
  \bibitem[Li et~al\mbox{.}(2022)]%
          {Li2022}
  \bibfield{author}{\bibinfo{person}{Yingwei Li}, \bibinfo{person}{Adams~Wei Yu}, \bibinfo{person}{Tianjian Meng}, \bibinfo{person}{Ben Caine}, \bibinfo{person}{Jiquan Ngiam}, \bibinfo{person}{Daiyi Peng}, \bibinfo{person}{Junyang Shen}, \bibinfo{person}{Yifeng Lu}, \bibinfo{person}{Denny Zhou}, \bibinfo{person}{Quoc~V Le}, {et~al\mbox{.}}} \bibinfo{year}{2022}\natexlab{}.
  \newblock \showarticletitle{Deepfusion: Lidar-camera deep fusion for multi-modal 3d object detection}.
  \newblock  (\bibinfo{year}{2022}), \bibinfo{pages}{17182--17191}.
  \newblock
  
  
  \bibitem[Liu et~al\mbox{.}(2022)]%
          {Liu2022}
  \bibfield{author}{\bibinfo{person}{Zixi Liu}, \bibinfo{person}{Yang Feng}, \bibinfo{person}{Yining Yin}, \bibinfo{person}{Jingyu Sun}, \bibinfo{person}{Zhenyu Chen}, {and} \bibinfo{person}{Baowen Xu}.} \bibinfo{year}{2022}\natexlab{}.
  \newblock \showarticletitle{QATest: A Uniform Fuzzing Framework for Question Answering Systems}.
  \newblock  (\bibinfo{year}{2022}), \bibinfo{pages}{1--12}.
  \newblock
  
  
  \bibitem[Ma et~al\mbox{.}(2018)]%
          {Ma2018}
  \bibfield{author}{\bibinfo{person}{Lei Ma}, \bibinfo{person}{Fuyuan Zhang}, \bibinfo{person}{Jiyuan Sun}, \bibinfo{person}{Minhui Xue}, \bibinfo{person}{Bo Li}, \bibinfo{person}{Felix Juefei-Xu}, \bibinfo{person}{Chao Xie}, \bibinfo{person}{Li Li}, \bibinfo{person}{Yang Liu}, \bibinfo{person}{Jianjun Zhao}, {et~al\mbox{.}}} \bibinfo{year}{2018}\natexlab{}.
  \newblock \showarticletitle{Deepmutation: Mutation testing of deep learning systems}.
  \newblock  (\bibinfo{year}{2018}), \bibinfo{pages}{100--111}.
  \newblock
  
  
  \bibitem[Midjourney(2023)]%
          {Midjourney}
  \bibfield{author}{\bibinfo{person}{Midjourney}.} \bibinfo{year}{2023}\natexlab{}.
  \newblock \bibinfo{title}{Midjourney}.
  \newblock \bibinfo{howpublished}{Website}.
  \newblock
  \newblock
  \shownote{\url{https://www.midjourney.com}}.
  
  
  \bibitem[OpenAI(2023)]%
          {DALLE}
  \bibfield{author}{\bibinfo{person}{OpenAI}.} \bibinfo{year}{2023}\natexlab{}.
  \newblock \bibinfo{title}{DALL$\cdot$E 2}.
  \newblock \bibinfo{howpublished}{Website}.
  \newblock
  \newblock
  \shownote{\url{https://openai.com/research/dall-e}}.
  
  
  \bibitem[PromptHero(2023)]%
          {Openjourney}
  \bibfield{author}{\bibinfo{person}{PromptHero}.} \bibinfo{year}{2023}\natexlab{}.
  \newblock \bibinfo{booktitle}{\emph{OpenJourney}}.
  \newblock
  \urldef\tempurl%
  \url{http://openjourney.art/}
  \showURL{%
  \tempurl}
  
  
  \bibitem[Qi et~al\mbox{.}(2020)]%
          {qi2020stanza}
  \bibfield{author}{\bibinfo{person}{Peng Qi}, \bibinfo{person}{Yuhao Zhang}, \bibinfo{person}{Yuhui Zhang}, \bibinfo{person}{Jason Bolton}, {and} \bibinfo{person}{Christopher~D. Manning}.} \bibinfo{year}{2020}\natexlab{}.
  \newblock \showarticletitle{Stanza: {A} Python Natural Language Processing Toolkit for Many Human Languages}. In \bibinfo{booktitle}{\emph{{ACL} (demo)}}. \bibinfo{publisher}{Association for Computational Linguistics}, \bibinfo{pages}{101--108}.
  \newblock
  
  
  \bibitem[Qiao et~al\mbox{.}(2019)]%
          {qiao2019mirrorgan}
  \bibfield{author}{\bibinfo{person}{Tingting Qiao}, \bibinfo{person}{Jing Zhang}, \bibinfo{person}{Duanqing Xu}, {and} \bibinfo{person}{Dacheng Tao}.} \bibinfo{year}{2019}\natexlab{}.
  \newblock \showarticletitle{Mirrorgan: Learning text-to-image generation by redescription}. In \bibinfo{booktitle}{\emph{Proceedings of the IEEE/CVF Conference on Computer Vision and Pattern Recognition}}. \bibinfo{pages}{1505--1514}.
  \newblock
  
  
  \bibitem[Radford et~al\mbox{.}(2021)]%
          {radford2021learning}
  \bibfield{author}{\bibinfo{person}{Alec Radford}, \bibinfo{person}{Jong~Wook Kim}, \bibinfo{person}{Chris Hallacy}, \bibinfo{person}{Aditya Ramesh}, \bibinfo{person}{Gabriel Goh}, \bibinfo{person}{Sandhini Agarwal}, \bibinfo{person}{Girish Sastry}, \bibinfo{person}{Amanda Askell}, \bibinfo{person}{Pamela Mishkin}, \bibinfo{person}{Jack Clark}, {et~al\mbox{.}}} \bibinfo{year}{2021}\natexlab{}.
  \newblock \showarticletitle{Learning transferable visual models from natural language supervision}. In \bibinfo{booktitle}{\emph{International conference on machine learning}}. PMLR, \bibinfo{pages}{8748--8763}.
  \newblock
  
  
  \bibitem[Ramesh et~al\mbox{.}(2021)]%
          {ramesh2021zero}
  \bibfield{author}{\bibinfo{person}{Aditya Ramesh}, \bibinfo{person}{Mikhail Pavlov}, \bibinfo{person}{Gabriel Goh}, \bibinfo{person}{Scott Gray}, \bibinfo{person}{Chelsea Voss}, \bibinfo{person}{Alec Radford}, \bibinfo{person}{Mark Chen}, {and} \bibinfo{person}{Ilya Sutskever}.} \bibinfo{year}{2021}\natexlab{}.
  \newblock \showarticletitle{Zero-shot text-to-image generation}. In \bibinfo{booktitle}{\emph{International Conference on Machine Learning}}. PMLR, \bibinfo{pages}{8821--8831}.
  \newblock
  
  
  \bibitem[Rombach et~al\mbox{.}(2022)]%
          {rombach2022}
  \bibfield{author}{\bibinfo{person}{Robin Rombach}, \bibinfo{person}{Andreas Blattmann}, \bibinfo{person}{Dominik Lorenz}, \bibinfo{person}{Patrick Esser}, {and} \bibinfo{person}{Bj{\"o}rn Ommer}.} \bibinfo{year}{2022}\natexlab{}.
  \newblock \showarticletitle{High-resolution image synthesis with latent diffusion models}. In \bibinfo{booktitle}{\emph{Proceedings of the IEEE/CVF conference on computer vision and pattern recognition}}. \bibinfo{pages}{10684--10695}.
  \newblock
  
  
  \bibitem[Saharia et~al\mbox{.}(2022)]%
          {Saharia2022}
  \bibfield{author}{\bibinfo{person}{Chitwan Saharia}, \bibinfo{person}{William Chan}, \bibinfo{person}{Saurabh Saxena}, \bibinfo{person}{Lala Li}, \bibinfo{person}{Jay Whang}, \bibinfo{person}{Emily~L Denton}, \bibinfo{person}{Kamyar Ghasemipour}, \bibinfo{person}{Raphael Gontijo~Lopes}, \bibinfo{person}{Burcu Karagol~Ayan}, \bibinfo{person}{Tim Salimans}, {et~al\mbox{.}}} \bibinfo{year}{2022}\natexlab{}.
  \newblock \showarticletitle{Photorealistic text-to-image diffusion models with deep language understanding}.
  \newblock \bibinfo{journal}{\emph{Advances in Neural Information Processing Systems}}  \bibinfo{volume}{35} (\bibinfo{year}{2022}), \bibinfo{pages}{36479--36494}.
  \newblock
  
  
  \bibitem[Salimans et~al\mbox{.}(2016)]%
          {salimans2016improved}
  \bibfield{author}{\bibinfo{person}{Tim Salimans}, \bibinfo{person}{Ian Goodfellow}, \bibinfo{person}{Wojciech Zaremba}, \bibinfo{person}{Vicki Cheung}, \bibinfo{person}{Alec Radford}, {and} \bibinfo{person}{Xi Chen}.} \bibinfo{year}{2016}\natexlab{}.
  \newblock \showarticletitle{Improved techniques for training gans}.
  \newblock \bibinfo{journal}{\emph{Advances in neural information processing systems}}  \bibinfo{volume}{29} (\bibinfo{year}{2016}), \bibinfo{pages}{1--10}.
  \newblock
  
  
  \bibitem[Saxon et~al\mbox{.}(2024)]%
          {saxon2024evaluates}
  \bibfield{author}{\bibinfo{person}{Michael Saxon}, \bibinfo{person}{Fatima Jahara}, \bibinfo{person}{Mahsa Khoshnoodi}, \bibinfo{person}{Yujie Lu}, \bibinfo{person}{Aditya Sharma}, {and} \bibinfo{person}{William~Yang Wang}.} \bibinfo{year}{2024}\natexlab{}.
  \newblock \showarticletitle{Who Evaluates the Evaluations? Objectively Scoring Text-to-Image Prompt Coherence Metrics with T2IScoreScore {(TS2)}}.
  \newblock \bibinfo{journal}{\emph{CoRR}}  \bibinfo{volume}{abs/2404.04251} (\bibinfo{year}{2024}).
  \newblock
  
  
  \bibitem[Stability.ai(2023)]%
          {Stability}
  \bibfield{author}{\bibinfo{person}{Stability.ai}.} \bibinfo{year}{2023}\natexlab{}.
  \newblock \bibinfo{title}{Stable Diffusion}.
  \newblock \bibinfo{howpublished}{Website}.
  \newblock
  \newblock
  \shownote{\url{https://stablediffusionweb.com}}.
  
  
  \bibitem[Sueyoshi and Matsubara(2024)]%
          {sueyoshi2024predicated}
  \bibfield{author}{\bibinfo{person}{Kota Sueyoshi} {and} \bibinfo{person}{Takashi Matsubara}.} \bibinfo{year}{2024}\natexlab{}.
  \newblock \showarticletitle{Predicated Diffusion: Predicate Logic-Based Attention Guidance for Text-to-Image Diffusion Models}. In \bibinfo{booktitle}{\emph{Proceedings of the IEEE/CVF Conference on Computer Vision and Pattern Recognition}}. \bibinfo{pages}{8651--8660}.
  \newblock
  
  
  \bibitem[Sun et~al\mbox{.}(2020)]%
          {sun2020automatic}
  \bibfield{author}{\bibinfo{person}{Zeyu Sun}, \bibinfo{person}{Jie~M Zhang}, \bibinfo{person}{Mark Harman}, \bibinfo{person}{Mike Papadakis}, {and} \bibinfo{person}{Lu Zhang}.} \bibinfo{year}{2020}\natexlab{}.
  \newblock \showarticletitle{Automatic testing and improvement of machine translation}. In \bibinfo{booktitle}{\emph{Proceedings of the ACM/IEEE 42nd international conference on software engineering}}. \bibinfo{pages}{974--985}.
  \newblock
  
  
  \bibitem[Szegedy et~al\mbox{.}(2016)]%
          {szegedy2016rethinking}
  \bibfield{author}{\bibinfo{person}{Christian Szegedy}, \bibinfo{person}{Vincent Vanhoucke}, \bibinfo{person}{Sergey Ioffe}, \bibinfo{person}{Jon Shlens}, {and} \bibinfo{person}{Zbigniew Wojna}.} \bibinfo{year}{2016}\natexlab{}.
  \newblock \showarticletitle{Rethinking the inception architecture for computer vision}. In \bibinfo{booktitle}{\emph{Proceedings of the IEEE conference on computer vision and pattern recognition}}. \bibinfo{pages}{2818--2826}.
  \newblock
  
  
  \bibitem[Tang et~al\mbox{.}(2020)]%
          {tang2020unbiased}
  \bibfield{author}{\bibinfo{person}{Kaihua Tang}, \bibinfo{person}{Yulei Niu}, \bibinfo{person}{Jianqiang Huang}, \bibinfo{person}{Jiaxin Shi}, {and} \bibinfo{person}{Hanwang Zhang}.} \bibinfo{year}{2020}\natexlab{}.
  \newblock \showarticletitle{Unbiased Scene Graph Generation from Biased Training}. In \bibinfo{booktitle}{\emph{Conference on Computer Vision and Pattern Recognition}}. \bibinfo{pages}{1--16}.
  \newblock
  
  
  \bibitem[Theis et~al\mbox{.}(2016)]%
          {theis2015note}
  \bibfield{author}{\bibinfo{person}{Lucas Theis}, \bibinfo{person}{A{\"{a}}ron van~den Oord}, {and} \bibinfo{person}{Matthias Bethge}.} \bibinfo{year}{2016}\natexlab{}.
  \newblock \showarticletitle{A note on the evaluation of generative models}. In \bibinfo{booktitle}{\emph{{ICLR}}}.
  \newblock
  
  
  \bibitem[Tian et~al\mbox{.}(2018)]%
          {Jana2018}
  \bibfield{author}{\bibinfo{person}{Yuchi Tian}, \bibinfo{person}{Kexin Pei}, \bibinfo{person}{Suman Jana}, {and} \bibinfo{person}{Baishakhi Ray}.} \bibinfo{year}{2018}\natexlab{}.
  \newblock \showarticletitle{Deeptest: Automated testing of deep-neural-network-driven autonomous cars}.
  \newblock  (\bibinfo{year}{2018}), \bibinfo{pages}{303--314}.
  \newblock
  
  
  \bibitem[Trabucco et~al\mbox{.}(2024)]%
          {dataaugmentation}
  \bibfield{author}{\bibinfo{person}{Brandon Trabucco}, \bibinfo{person}{Kyle Doherty}, \bibinfo{person}{Max Gurinas}, {and} \bibinfo{person}{Ruslan Salakhutdinov}.} \bibinfo{year}{2024}\natexlab{}.
  \newblock \showarticletitle{Effective Data Augmentation With Diffusion Models}. In \bibinfo{booktitle}{\emph{{ICLR}}}. \bibinfo{publisher}{OpenReview.net}.
  \newblock
  
  
  \bibitem[Tredinnick and Laybats(2023)]%
          {dangers}
  \bibfield{author}{\bibinfo{person}{Luke Tredinnick} {and} \bibinfo{person}{Claire Laybats}.} \bibinfo{year}{2023}\natexlab{}.
  \newblock \showarticletitle{The dangers of generative artificial intelligence}.
  \newblock \bibinfo{journal}{\emph{Business Information Review}}  \bibinfo{volume}{40} (\bibinfo{year}{2023}), \bibinfo{pages}{02663821231183756}.
  \newblock
  
  
  \bibitem[Wan et~al\mbox{.}(2024)]%
          {wan2024factuality}
  \bibfield{author}{\bibinfo{person}{Yixin Wan}, \bibinfo{person}{Di Wu}, \bibinfo{person}{Haoran Wang}, {and} \bibinfo{person}{Kai{-}Wei Chang}.} \bibinfo{year}{2024}\natexlab{}.
  \newblock \showarticletitle{The Factuality Tax of Diversity-Intervened Text-to-Image Generation: Benchmark and Fact-Augmented Intervention}.
  \newblock \bibinfo{journal}{\emph{CoRR}}  \bibinfo{volume}{abs/2407.00377} (\bibinfo{year}{2024}).
  \newblock
  
  
  \bibitem[Wu et~al\mbox{.}(2022)]%
          {9879249}
  \bibfield{author}{\bibinfo{person}{Fuxiang Wu}, \bibinfo{person}{Liu Liu}, \bibinfo{person}{Fusheng Hao}, \bibinfo{person}{Fengxiang He}, {and} \bibinfo{person}{Jun Cheng}.} \bibinfo{year}{2022}\natexlab{}.
  \newblock \showarticletitle{Text-to-Image Synthesis based on Object-Guided Joint-Decoding Transformer}. In \bibinfo{booktitle}{\emph{2022 IEEE/CVF Conference on Computer Vision and Pattern Recognition (CVPR)}}. \bibinfo{pages}{18092--18101}.
  \newblock
  \urldef\tempurl%
  \url{https://doi.org/10.1109/CVPR52688.2022.01758}
  \showDOI{\tempurl}
  
  
  \bibitem[Wu(2022)]%
          {Art1}
  \bibfield{author}{\bibinfo{person}{Xianchao Wu}.} \bibinfo{year}{2022}\natexlab{}.
  \newblock \showarticletitle{Creative Painting with Latent Diffusion Models}.
  \newblock \bibinfo{journal}{\emph{CoRR}}  \bibinfo{volume}{abs/2209.14697} (\bibinfo{year}{2022}).
  \newblock
  
  
  \bibitem[Xie et~al\mbox{.}(2023)]%
          {xie2023qaasker+}
  \bibfield{author}{\bibinfo{person}{Xiaoyuan Xie}, \bibinfo{person}{Shuo Jin}, {and} \bibinfo{person}{Songqiang Chen}.} \bibinfo{year}{2023}\natexlab{}.
  \newblock \showarticletitle{qaAskeR+: a novel testing method for question answering software via asking recursive questions}.
  \newblock \bibinfo{journal}{\emph{Automated Software Engineering}} \bibinfo{volume}{30}, \bibinfo{number}{1} (\bibinfo{year}{2023}), \bibinfo{pages}{14}.
  \newblock
  
  
  \bibitem[Xie et~al\mbox{.}(2024)]%
          {xie2024metamorphic}
  \bibfield{author}{\bibinfo{person}{Xiaoyuan Xie}, \bibinfo{person}{Xingpeng Li}, {and} \bibinfo{person}{Songqiang Chen}.} \bibinfo{year}{2024}\natexlab{}.
  \newblock \showarticletitle{Metamorphic Testing of Image Captioning Systems via Image-Level Reduction}.
  \newblock \bibinfo{journal}{\emph{IEEE Transactions on Software Engineering}} (\bibinfo{year}{2024}).
  \newblock
  
  
  \bibitem[Xie et~al\mbox{.}(2020)]%
          {xie2020mettle}
  \bibfield{author}{\bibinfo{person}{Xiaoyuan Xie}, \bibinfo{person}{Zhiyi Zhang}, \bibinfo{person}{Tsong~Yueh Chen}, \bibinfo{person}{Yang Liu}, \bibinfo{person}{Pak-Lok Poon}, {and} \bibinfo{person}{Baowen Xu}.} \bibinfo{year}{2020}\natexlab{}.
  \newblock \showarticletitle{METTLE: A metamorphic testing approach to assessing and validating unsupervised machine learning systems}.
  \newblock \bibinfo{journal}{\emph{IEEE Transactions on Reliability}} \bibinfo{volume}{69}, \bibinfo{number}{4} (\bibinfo{year}{2020}), \bibinfo{pages}{1293--1322}.
  \newblock
  
  
  \bibitem[Xu et~al\mbox{.}(2018)]%
          {xu2018attngan}
  \bibfield{author}{\bibinfo{person}{Tao Xu}, \bibinfo{person}{Pengchuan Zhang}, \bibinfo{person}{Qiuyuan Huang}, \bibinfo{person}{Han Zhang}, \bibinfo{person}{Zhe Gan}, \bibinfo{person}{Xiaolei Huang}, {and} \bibinfo{person}{Xiaodong He}.} \bibinfo{year}{2018}\natexlab{}.
  \newblock \showarticletitle{Attngan: Fine-grained text to image generation with attentional generative adversarial networks}. In \bibinfo{booktitle}{\emph{Proceedings of the IEEE conference on computer vision and pattern recognition}}. \bibinfo{pages}{1316--1324}.
  \newblock
  
  
  \bibitem[Yu et~al\mbox{.}(2022b)]%
          {Yu2022}
  \bibfield{author}{\bibinfo{person}{Boxi Yu}, \bibinfo{person}{Zhiqing Zhong}, \bibinfo{person}{Xinran Qin}, \bibinfo{person}{Jiayi Yao}, \bibinfo{person}{Yuancheng Wang}, {and} \bibinfo{person}{Pinjia He}.} \bibinfo{year}{2022}\natexlab{b}.
  \newblock \showarticletitle{Automated testing of image captioning systems}. In \bibinfo{booktitle}{\emph{Proceedings of the 31st ACM SIGSOFT International Symposium on Software Testing and Analysis}}. \bibinfo{pages}{467--479}.
  \newblock
  
  
  \bibitem[Yu et~al\mbox{.}(2022a)]%
          {yu2022scaling}
  \bibfield{author}{\bibinfo{person}{Jiahui Yu}, \bibinfo{person}{Yuanzhong Xu}, \bibinfo{person}{Jing~Yu Koh}, \bibinfo{person}{Thang Luong}, \bibinfo{person}{Gunjan Baid}, \bibinfo{person}{Zirui Wang}, \bibinfo{person}{Vijay Vasudevan}, \bibinfo{person}{Alexander Ku}, \bibinfo{person}{Yinfei Yang}, \bibinfo{person}{Burcu~Karagol Ayan}, \bibinfo{person}{Ben Hutchinson}, \bibinfo{person}{Wei Han}, \bibinfo{person}{Zarana Parekh}, \bibinfo{person}{Xin Li}, \bibinfo{person}{Han Zhang}, \bibinfo{person}{Jason Baldridge}, {and} \bibinfo{person}{Yonghui Wu}.} \bibinfo{year}{2022}\natexlab{a}.
  \newblock \showarticletitle{Scaling Autoregressive Models for Content-Rich Text-to-Image Generation}.
  \newblock \bibinfo{journal}{\emph{Trans. Mach. Learn. Res.}}  \bibinfo{volume}{2022} (\bibinfo{year}{2022}).
  \newblock
  
  
  \bibitem[Yu et~al\mbox{.}(2023)]%
          {yu2023visually}
  \bibfield{author}{\bibinfo{person}{Qifan Yu}, \bibinfo{person}{Juncheng Li}, \bibinfo{person}{Yu Wu}, \bibinfo{person}{Siliang Tang}, \bibinfo{person}{Wei Ji}, {and} \bibinfo{person}{Yueting Zhuang}.} \bibinfo{year}{2023}\natexlab{}.
  \newblock \showarticletitle{Visually-Prompted Language Model for Fine-Grained Scene Graph Generation in an Open World}. In \bibinfo{booktitle}{\emph{{ICCV}}}. \bibinfo{publisher}{{IEEE}}, \bibinfo{pages}{21503--21514}.
  \newblock
  
  
  \bibitem[Zhang et~al\mbox{.}(2017)]%
          {zhang2017stackgan}
  \bibfield{author}{\bibinfo{person}{Han Zhang}, \bibinfo{person}{Tao Xu}, \bibinfo{person}{Hongsheng Li}, \bibinfo{person}{Shaoting Zhang}, \bibinfo{person}{Xiaogang Wang}, \bibinfo{person}{Xiaolei Huang}, {and} \bibinfo{person}{Dimitris~N Metaxas}.} \bibinfo{year}{2017}\natexlab{}.
  \newblock \showarticletitle{Stackgan: Text to photo-realistic image synthesis with stacked generative adversarial networks}. In \bibinfo{booktitle}{\emph{Proceedings of the IEEE international conference on computer vision}}. \bibinfo{pages}{5907--5915}.
  \newblock
  
  
  \bibitem[Zhang et~al\mbox{.}(2024)]%
          {zhang2024machine}
  \bibfield{author}{\bibinfo{person}{Quanjun Zhang}, \bibinfo{person}{Juan Zhai}, \bibinfo{person}{Chunrong Fang}, \bibinfo{person}{Jiawei Liu}, \bibinfo{person}{Weisong Sun}, \bibinfo{person}{Haichuan Hu}, {and} \bibinfo{person}{Qingyu Wang}.} \bibinfo{year}{2024}\natexlab{}.
  \newblock \showarticletitle{Machine Translation Testing via Syntactic Tree Pruning}.
  \newblock \bibinfo{journal}{\emph{ACM Transactions on Software Engineering and Methodology}} \bibinfo{volume}{33}, \bibinfo{number}{5} (\bibinfo{year}{2024}), \bibinfo{pages}{1--39}.
  \newblock
  
  
  \bibitem[Zhao et~al\mbox{.}(2024)]%
          {zhao2024flasheval}
  \bibfield{author}{\bibinfo{person}{Lin Zhao}, \bibinfo{person}{Tianchen Zhao}, \bibinfo{person}{Zinan Lin}, \bibinfo{person}{Xuefei Ning}, \bibinfo{person}{Guohao Dai}, \bibinfo{person}{Huazhong Yang}, {and} \bibinfo{person}{Yu Wang}.} \bibinfo{year}{2024}\natexlab{}.
  \newblock \showarticletitle{FlashEval: Towards Fast and Accurate Evaluation of Text-to-image Diffusion Generative Models}. In \bibinfo{booktitle}{\emph{Proceedings of the IEEE/CVF Conference on Computer Vision and Pattern Recognition}}. \bibinfo{pages}{16122--16131}.
  \newblock
  
  
  \bibitem[Zhu et~al\mbox{.}(2017)]%
          {zhu2017unpaired}
  \bibfield{author}{\bibinfo{person}{Jun-Yan Zhu}, \bibinfo{person}{Taesung Park}, \bibinfo{person}{Phillip Isola}, {and} \bibinfo{person}{Alexei~A Efros}.} \bibinfo{year}{2017}\natexlab{}.
  \newblock \showarticletitle{Unpaired image-to-image translation using cycle-consistent adversarial networks}. In \bibinfo{booktitle}{\emph{Proceedings of the IEEE international conference on computer vision}}. \bibinfo{pages}{2223--2232}.
  \newblock
  
  
  \end{thebibliography}

\end{document}